\documentclass{article}

\usepackage{arxiv}

\usepackage[utf8]{inputenc} 
\usepackage[T1]{fontenc}    
\usepackage[colorlinks=true,citecolor=blue, linkcolor=blue]{hyperref}       
\usepackage{url}            
\usepackage{booktabs}       
\usepackage{amsfonts}       
\usepackage{nicefrac}       
\usepackage{microtype}      
\usepackage{lipsum}         
\usepackage{graphicx}
\usepackage[numbers]{natbib}
\usepackage{doi}

\usepackage{amssymb}
\usepackage{listings}
\usepackage{float}
\usepackage{lineno}
\usepackage{siunitx}
\usepackage{graphicx,color,colordvi}
\usepackage{latexsym,amsbsy,epsfig,fancybox}
\usepackage{amstext}
\usepackage{subcaption}
\usepackage{amsfonts,textgreek}
\usepackage{mathrsfs}
\usepackage{mathtools}
\usepackage{url}
\usepackage{wrapfig,bm}
\usepackage{tikz}
\usepackage{pstricks}
\usepackage{caption}
\usepackage{psfrag}
\usepackage{latexsym}
\usepackage{array}
\usepackage{multirow}
\usepackage{booktabs}
\usepackage{colortbl}
\usepackage{verbatim}
\usepackage{color}
\usepackage{epstopdf,overpic}
\usepackage{enumerate}
\usepackage[nameinlink]{cleveref}
\usepackage{amsmath,amssymb,xspace}
\usetikzlibrary{plotmarks}
\usetikzlibrary{fit,positioning,arrows}
\usetikzlibrary{shapes.geometric, plotmarks}
\usepackage{appendix}

\usepackage{longtable}
\usepackage{lipsum}
\usepackage{multicol}
\usepackage{setspace}   
\usepackage{bbm}

\bgroup\catcode`\#=12\gdef\hash{#}\egroup 
\newcommand{\bkt}[1]{\left(#1\right)}
\tikzstyle{nicebox}=[draw=black!100, fill=white!10, rectangle, inner sep=4pt, inner ysep=16pt]
\tikzstyle{niceboxtitle}=[draw=black!100, fill=white, text=black, rectangle]

\usetikzlibrary{arrows,decorations.markings}

\title{Physics-informed neural networks for solving thermo-mechanics problems of functionally graded material}

\author{\href{https://orcid.org/0000-0002-0538-9479}{\includegraphics[scale=0.06]{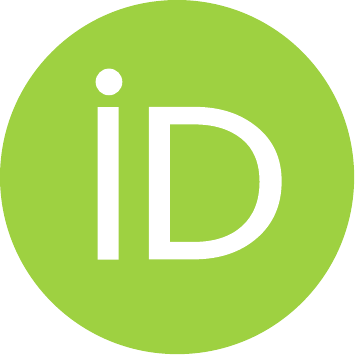}\hspace{1mm}Mayank Raj} \\
	Department of Mechanical Engineering\\
	Indian Institute of Technology Madras\\
	Chennai-600036\\
	\texttt{me16b147@smail.iitm.ac.in} \\
	\And
	\href{https://orcid.org/0000-0003-4486-8680}{\includegraphics[scale=0.06]{orcid.pdf}\hspace{1mm}Pramod Kumbhar} \\
	Department of Mechanical Engineering\\
	Indian Institute of Technology Madras\\
	Chennai-600036\\
	\texttt{pramodkumbhar789@gmail.com} \\
		\And
	\href{https://orcid.org/
0000-0002-9492-8592}{\includegraphics[scale=0.06]{orcid.pdf}\hspace{1mm}Ratna Kumar Annabattula}\thanks{Corresponding author} \\
	Department of Mechanical Engineering\\
	Indian Institute of Technology Madras\\
	Chennai-600036\\
	\texttt{ratna@iitm.ac.in} \\
}

\renewcommand{\shorttitle}{Physics-informed neural network}

\begin{document}
\maketitle

\begin{abstract}
	Differential equations are indispensable to engineering and hence to innovation. In recent years, physics-informed neural networks (PINN) have emerged as a novel method for solving differential equations. PINN method has the advantage of being meshless, scalable, and can potentially be intelligent in terms of transferring the knowledge learned from solving one differential equation to the other. The exploration in this field has majorly been limited to solving linear-elasticity problems, crack propagation problems. This study uses PINNs to solve coupled thermo-mechanics problems of materials with functionally graded properties.  An in-depth analysis of the PINN framework has been carried out by understanding the training datasets, model architecture, and loss functions. The efficacy of the PINN models in solving thermo-mechanics differential equations has been measured by comparing the obtained solutions either with analytical solutions or finite element method-based solutions. While R\textsuperscript{2} score of more than 99\% has been achieved in predicting primary variables such as displacement and temperature fields, achieving the same for secondary variables such as stress turns out to be more challenging. This study is the first to implement the PINN framework for solving coupled thermo-mechanics problems on composite materials. This study is expected to enhance the understanding of the novel PINN framework and will be seminal for further research on PINNs.
\end{abstract}


\section{Introduction}
The term ``neural network" encompasses a wide range of machine learning models with one similarity, i.e., so-called ``neurons" as the basic building block \cite{Goodfellow2016, LeCun2015}. Neural networks have seen many applications in fields not limited to computer vision \cite{Huang2020CV}, banking and finance \cite{Huang2020}, bio-informatics \cite{Tang2019, ardabili2020covid}, renewable energy systems \cite{peeketi2019thermal, mosavi2019state} and material science \cite{Masubuchi2020, Raj2021, liu2021stochastic}. In other words, if there is a function approximation task, neural networks are indisputable models to use. The use cases are not limited to just function approximation tasks. Rather, neural networks dominate in image captioning and language translation \cite{Stephen2017}, which are one-to-many or many-to-many mapping tasks. So what is behind the success of neural networks? Certainly, the flexibility they offer in designing composite parametric functions and the ease of optimizing those parameters \cite{Rumelhart1986} lends to their success. Often called universal function approximators \cite{Hornik1989}, these neural networks can, in theory, fit most of the relational mappings. However, the challenge lies in designing a neural network architecture so that at least one set of optimum weights fits the desired function. The so-called training of a neural network, i.e., estimation of optimal parameters using an optimization algorithm, can also be challenging. However, after going through those steps, one gets a trained neural network model that can predict/estimate the desired target variable on a given set of inputs in seconds. Alongside all the success stories, neural networks also have their own set of drawbacks. Training often requires considerable data, careful data sampling to cover variance in the domain and target space, difficult parallelization of the training process, and dedicated hardware such as GPUs \cite{Oh2014} for fast training.

Differential equations are indispensable to computational engineering. Irrespective of engineering subdomains, differential equations appear in one form or another. Often one needs to solve such equations on the domain of interest numerically. The state of the art of technique for numerically solving such differential equations is the finite element method. While the finite element method has its pros and cons, the physics-informed neural network (PINN) is a promising alternative to solving differential equations. 

The family of a neural network is vast, and variety can be in various forms, such as their architecture (arrangement of neurons and layers), learning process (supervised or unsupervised), the purpose of the model (regression or classification). Physics-informed neural network (PINN) is yet another variant in this family. Presently in its infancy, PINN can potentially disrupt how differential equations are solved. PINN has an apparent advantage of being a meshless method over FEM. Unlike conventional supervised learning Deep Neural Networks (DNNs), PINN models are trained without the target samples. While PINN in its current form can take more time than commercial finite element solvers such as Abaqus\textsuperscript{\textregistered} to solve simple elasticity problems, it can do better when simulations get complex and the number of simulations increases, such as in an optimization scenario. PINN can use the neural network transfer learning concept \cite{Tan2018} to train on new samples in marginal time.

In an abstract sense, a physics-informed neural network tries to optimize the parameters of a parametric function that approximates the solution of the differential equation on a given domain. The network does the optimization by minimizing an objective called the loss function in the machine learning community. The prefix ``Physics-informed" comes from the fact that the loss function is obtained using the underlying physics of the problem represented by the differential equations. Often, the loss is the functional corresponding to the differential equation. 

The physics-informed neural network was first introduced by Raissi et al. \cite{RAISSI2019} as an alternate method to solve partial differential equations. PINNs can be classified into two categories based on the implementation of the loss function. One of the approaches uses the collocation technique, where simply the residual of the differential equation is minimized by training the neural network \cite{Weinan2018, anitescu2019artificial}. While on the other hand, there are energy-based loss functions where the functional of the differential equation is minimized \cite{samaniego2020energy}. While the collocation approach is straightforward, it involves computing higher-order derivatives, resulting in difficult training.
On the other hand, the energy-based method \cite{samaniego2020energy} requires lower order derivatives due to the functional's weak form of the differential equation. Hence, energy-based methods have become more popular with numerous applications in the last few years. It is used in the prediction of crack propagation using the phase-field method \cite{goswami2020transfer, Goswami2020}. Further, the ease with which adaptive meshing can be done in PINN was shown by Goswami et al. \cite{goswami2020adaptive}. Li et al. \cite{li2021physics} compared the effectiveness of the PINN approach with other ML-based techniques in terms of various metrics such as coefficient of determination. Recently, Dwivedi et al. \cite{Dwivedi2020-1, DWIVEDI2020} introduced Physics Informed Extreme Learning Machine, which improves the efficiency of Physics Informed Neural Network by utilizing ideas of Extreme Learning Machine. Dwivedi et al. also introduced Distributed Physics Informed Neural Network \cite{dwivedi2019distributed, DWIVEDI2021, yadav2020} where the domain can be split into multiple sub-domains, and separate PINNs can be trained over them with ensuring continuity of variables on the interface of sub-domains.

Functionally graded materials (FGMs) are a unique class of composites incorporating the material properties of two or more different homogeneous materials~\cite{koizumi1993concept}. The material property variation is usually continuous and smooth such that it can be represented as explicit smooth mathematical functions.
Usually, these materials trump the conventional homogeneous components in terms of superior thermal and mechanical capabilities. Hence, it is not surprising that the FGMs find ubiquitous applications ranging from mechanical to bio-medical applications~\cite{pindera1997use}.

While there are many papers published in PINNs in the last year, none of them presents a comprehensive discussion on the pros and cons of the methodology. Moreover, the application has been limited to isotropic material, and discussions have mostly been centered around the primary variables such as displacement fields and the damage parameter in phase-field modeling of fracture. There has not been any discussion of secondary variables such as strain and stress fields that depend on primary variables' derivatives. Moreover, there has been a knowledge gap in understanding the convergence of energy-based PINN neural networks.

This paper starts with a discussion on using PINN models to solve one-dimensional differential equations. We specifically solve coupled thermo-mechanics problems in solid mechanics for the first time using PINN. We consider materials with spatially varying properties and show the robustness of PINN to arrive at the desired solution. We further discuss how and how not to implement different kinds of boundary conditions. Further on, we point out that the PINN model converges to a physically meaningful loss function value. We then demonstrate the efficacy of PINN in solving the two-dimensional thermo-mechanics problem with functionally graded properties. We achieve remarkable PINN prediction accuracy of displacement and temperature fields. However, we show that excellent accuracy in primary variables such as displacement fields does not imply the same for secondary variables such as stress fields. We conclude the paper by discussing further avenues of research on PINN models.

\section{Methods}
\label{sec:methods}
\subsection{Governing differential equations for thermo-elasticity}
\label{sec:governing_equations}
\begin{figure}[h]
    \centering
    \includegraphics[width=0.6\textwidth]{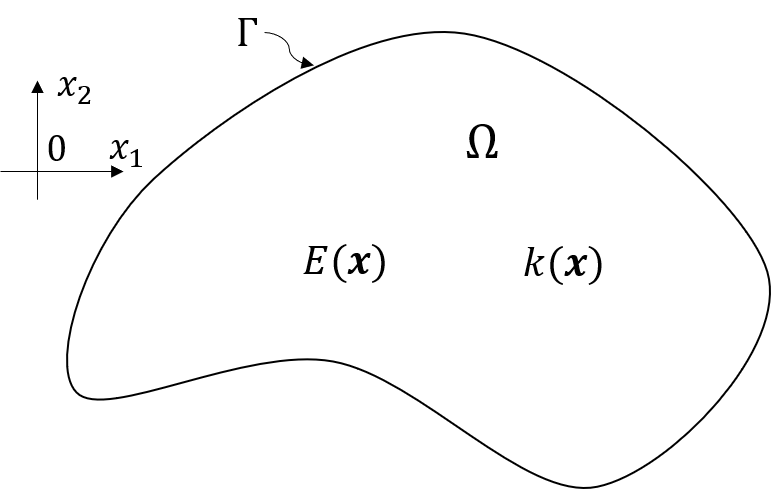}
    \caption{Schematic of a solid object with spatially varying properties lying in a two-dimensional space.}
    \label{fig:body_schematic}
\end{figure}
Let us consider a solid body occupying spatial domain $\Omega \subset \mathcal{R}^d$. The solid body is composed of material with thermo-elastic properties (see~\Cref{fig:body_schematic}). The $d$ dimensional domain is bounded by a surface $\Gamma$. The bounding surface is defined locally in terms of the outward unit normal vector $\mathbf{n}$. Let the elastic modulus and thermal conductivity of body vary as $E(\mathbf{x})$ and $k(\mathbf{x})$ where $\mathbf{x}$ is a point in the domain $\Omega$. The static thermo-elastic response of the body, displacement field $\mathbf{u}: \Omega \to R^d$ and temperature field $T:\Omega \to R$ is governed by the following differential equations:

\begin{equation}
    \nabla \cdot \mathbf{\sigma}(\mathbf{x}) + \mathbf{f}(\mathbf{x}) = 0\quad \forall \mathbf{x} \in \Omega
    \label{eq:solid_mechanics_equation}
\end{equation}

\begin{equation}
    \nabla \cdot \bkt{k(\mathbf{x}) \nabla T(\mathbf{x})} = 0 \quad \forall \mathbf{x} \in \Omega
    \label{eq:thermal_equation}
\end{equation}

under the boundary conditions:

\begin{equation}
\begin{split}
    \mathbf{u}(\mathbf{x}) &= \mathbf{u_{bc}(x)}\quad \forall \mathbf{x} \in \Gamma_u\\
    \mathbf{\sigma(x)}\cdot \mathbf{n} &= \mathbf{t_{bc}(x)}\quad \forall \mathbf{x} \in \Gamma_t\\
    T(\mathbf{x}) &= T_{bc}(\mathbf{x}) \quad \forall \mathbf{x} \in \Gamma_T
\end{split}
    \label{eq:boundary_condition}
\end{equation}
\noindent $\mathbf{\sigma(x)}$ and $\mathbf{f(x)}$ represents stress field and body force respectively. Moreover, the prescribed boundary conditions $\mathbf{u_{bc}}$, $\mathbf{t_{bc}}$ and $T_{bc}$ are defined such that $\bkt{\Gamma_u \cup \Gamma_t \cup \Gamma_T} \subset \Gamma$ and $\Gamma_u \cap \Gamma_t = \Phi$. The stress field is related to strain field which in turn is related to displacement and temperature fields as follows:

\begin{equation}
    \begin{split}
    \mathbf{\sigma(x)} &= \mathbf{C(x)} : \mathbf{\varepsilon(x)}\\
    \mathbf{\varepsilon(x)} &=   \mathbf{\varepsilon}^{el}(\mathbf{x}) - \alpha \Delta T(\mathbf{x})\mathbf{I}\\
    \varepsilon^{el}_{ij} &= \dfrac{u_{i,j} + u_{j,i}}{2}
    \end{split}
    \label{eq:strain}
\end{equation}
\noindent where $\mathbf{\varepsilon}^{el}$ is the elastic strain while $\alpha \Delta T$ is the thermal strain. $\alpha$ is the coefficient of thermal expansion , $\mathbf{I}$ is the identity matrix and $\Delta T(\mathbf{x}) = T(\mathbf{x}) - T_0$ where $T_0$ is the temperature of the body in the reference state. The elastic material constitutive matrix $\mathbf{C(x)}$ is written in terms of elastic modulus and Poisson's ratio $\nu$ as:
\begin{equation}\mathbf{C(x)} = E(\mathbf{x})\begin{bmatrix} \dfrac{1}{1-\nu^2} & \dfrac{\nu}{1-\nu^2} & 0 \\ \dfrac{\nu}{1-\nu^2} & \dfrac{1}{1-\nu^2} & 0 \\ 0 & 0 & \dfrac{1}{(1+\nu)}\end{bmatrix}
\label{eq:stiffness_matrix}
\end{equation}
\noindent for two-dimensional object with plane-stress assumption.

The functional corresponding to differential equations governing the thermo-elastic response of solid body is given as:
 \begin{equation}
    \begin{split}
    \mathcal{F} &= W_{\text{int}} - W_{\text{ext} }\\
    &= \int_{\Omega} \frac{1}{2} \mathbf{\varepsilon}^{el}(\mathbf{x}):\mathbf{C(x)} : \mathbf{\varepsilon}^{el}(\mathbf{x}) d \Omega + \int_{\Omega} \frac{1}{2} k(\mathbf{x})\nabla T(\mathbf{x}) \cdot \nabla T(\mathbf{x}) d \Omega - \bkt{\int_{\Omega} \mathbf{f(x)} \cdot \mathbf{n} \mkern3mu d \Omega + \int_{d \Gamma_t} \mathbf{t(x)} \cdot \mathbf{n} \mkern3mu d\Gamma}
    \end{split}
    \label{eq:functional}
\end{equation}

The functional $\mathcal{F}$ can be viewed as internal energy $(W_{\text{int}})$ less work done by the external forces $(W_{\text{ext}})$.

In this study, there are a few examples where we have used PINN to predict only the elastic response of the object under consideration. In that scenario, only \Cref{eq:solid_mechanics_equation} governs the response of object. Moreover, the coefficient of thermal expansion $\bkt{\alpha}$ is set to be zero, and the internal energy corresponding to temperature gradient is removed from \Cref{eq:thermal_equation}. After setting up the differential equations and boundary conditions governing the problem at hand, PINN is used to minimize the functional, thereby obtaining the desired response fields $\mathbf{u(x)}$ and $T(\mathbf{x})$. The details of the PINN framework are discussed in the following subsection.

\subsection{Physics informed neural network}
\label{sec:physics_informed_neural_network}
Neural networks are composite parametric functions designed to learn any relational mapping. While the traditional neural networks used to have only one hidden layer, the abundance of computational resources has made it possible to design and train models with numerous hidden layers. Hence the term \textit{deep} neural network (DNN) has become synonymous with neural networks. The process of training a DNN is called deep learning. A DNN is built up using the following building blocks: (1) neurons, (2) input layer, (3) output layer, (4) hidden layers, (5) activation  function/s, and (6) weights and biases. Mathematically, a fully-connected deep neural network with straight-forward architecture can be described as:
\begin{align}
    \begin{split}
    \mathbf{h}^1 &= \psi(\mathbf{W}^1\mathbf{x} + \mathbf{b}^1)\\
    \mathbf{h}^2 &= \psi(\mathbf{W}^1\mathbf{h}^1 + \mathbf{b}^1)\\
    \vdots\\
    \mathbf{h}^{i} &= \psi(\mathbf{W}^i\mathbf{h}^{i-1} + \mathbf{b}^i)\\
    \vdots\\
    \mathbf{\tilde{v}} &= \mathbf{W}^o\mathbf{h}^{o-1} + \mathbf{b}^o
    \end{split}
    \label{eq:nn}
\end{align}
\noindent where $\mathbf{x}$ and $\mathbf{\tilde{v}}$ are respectively the input and output vectors. $\mathbf{W}^i$ are the weight matrices, $\mathbf{b}^i$ are the bias parameters,  $\mathbf{h}^i$ are hidden layer vectors and  $\psi(.)$ is the activation function used in different layers. Moreover, a further transformation, $\zeta:\mathcal{R}^m \to \mathcal{R}^m$ is applied on $\tilde{\mathbf{v}}$ in order to satisfy Dirichlet boundary conditions:
\begin{equation}
    \mathbf{\tilde{z}}(\mathbf{x}) =\zeta \bkt{\mathbf{x},\mathbf{\tilde{v}(x)}}
    \label{eq:dirichlet}
\end{equation}
More on the transformation $\zeta(.)$ is discussed in \Cref{sec:results_and_discussion}. Hence, the sequence of operations in \Cref{eq:nn,eq:dirichlet} can be represented as a DNN $  \mathbf{\tilde{z}}(\mathbf{x}):\mathcal{R}^m \to \mathcal{R}^n
$, which maps an $m$-dimensional real vector $\mathbf{x}$ to $n$-dimensional real vector $\mathbf{\tilde{z}}$. In the Results and Discussion \Cref{sec:results_and_discussion} $\mathbf{\tilde{u}}$ and ${\tilde{T}}$ will be often used in place of $\mathbf{\tilde{z}}$. The vector $\mathbf{\tilde{z}}$ will In order to optimize parameters of a conventional DNN, a loss function is defined as $\mathcal{L}(\mathbf{\tilde{z}}, \mathbf{z}): \mathcal{R}^{m\times m} \to \mathcal{R}
$, where $\mathbf{z}$ is true value of the target variable and $\mathcal{L}(.)$ can represent a loss function such as mean squared error. However, in case of PINN, the loss function is defined only in terms of $\mathbf{\tilde{u}}$ and its derivatives with respect to $\mathbf{x}$, for example: $\mathcal{L}(\mathbf{\tilde{z}}, \mathbf{\tilde{z}_x}): \mathcal{R}^{m\times m \times n} \to \mathcal{R}$.

In order to solve the ODEs given in \Cref{eq:solid_mechanics_equation,eq:thermal_equation} under the boundary conditions in \Cref{eq:boundary_condition}, the functional given is \Cref{eq:functional} must be discretized to obtain the loss function $\mathcal{L}$ for training the PINN. Hence, the discretized functional can be written as:

\begin{equation}
    \begin{split}
    \mathcal{L}&= \sum_{\mathbf{x}_i \in \Omega} \frac{1}{2} w(\mathbf{x}_i)\mathbf{\tilde{\varepsilon}}^{el}(\mathbf{x}_i):\mathbf{C}(\mathbf{x}_i) : \mathbf{\tilde{\varepsilon}}^{el}(\mathbf{x}_i)\\  &+ \sum_{\mathbf{x}_i \in \Omega} \frac{1}{2} w(\mathbf{x}_i)k(\mathbf{x}_i)\nabla \tilde{T}(\mathbf{x}_i) \cdot \nabla \tilde{T}(\mathbf{x}_i)\\ &- \bkt{\sum_{\mathbf{x}_i \in \Omega}w(\mathbf{x}_i) \mathbf{f(\mathbf{x}_i)} \cdot \mathbf{n} + \sum_{\mathbf{x}_i \in d \Gamma_t} w'(\mathbf{x}_i)\mathbf{t}(\mathbf{x}_i) \cdot \mathbf{n}}
    \end{split}
    \label{eq:loss_function}
\end{equation}
\noindent where $w(\mathbf{x})$ and $w'(\mathbf{x})$ represent the weights for numerical integration on domain and boundary nodes respectively.
The PINN predicted solutions $\mathbf{\hat{u}(x)}$ and $\hat{T}(\mathbf{x})$ is obtained by minimizing $\mathcal{L}$ such that:

\begin{equation}
    \begin{bmatrix}
    \mathbf{\hat{u}(x)}\\ {\hat{T}\mathbf{(x)}}\    \end{bmatrix} = \mathbf{\tilde{z}}= \arg \inf_{\tilde{\mathbf{z}}}\mathcal{L}(\tilde{\mathbf{z}},\tilde{\mathbf{z}}_x)
\end{equation}

\subsection{Training physics informed neural network}
\label{sec:training_neural_network}
The parameters of DNN: weights and biases are optimized using the backpropagation \cite{Rumelhart1986} algorithm. The algorithm computes the derivative of the loss function with respect to all parameters and subsequently updates them using the gradient descent rule. The weight update process is repeated for a number of epochs until the convergence criteria are satisfied. The parameters of the PINN model are updated using the Adam optimizer \cite{kingma2017adam}, which is an enhanced version of gradient descent. Computing derivatives in the backpropagation algorithm is at the core of the training process. The technique called automatic differentiation \cite{Baydin2018} is used for computing derivatives and hence is at the core of any deep learning library such as Tensorflow \cite{tensorflow2015-whitepaper}. To understand automatic differentiation, let's take a step back to reflect on how neural networks (or any such parametric function) are built in Tensorflow. To build a neural network, one uses the so-called building block parametric operators in Tensorflow. Examples of such operators are nothing but matrix multiplication,  addition, and an activation function. It is to be noted that the gradients to all such operators are already defined in Tensorflow. For example, let's look at the matrix multiplication operation. Let us say there is a matrix $\mathbf{W}$, which when multiplied with vector $\mathbf{x}$ of compatible size, returns the matrix multiplication result from $\mathbf{y} = \mathbf{W}\mathbf{x}$. Hence, whenever one needs to compute the gradient of this operation i.e, ${dy_i}/{dx_j}$, the result is the matrix $\mathbf{W}$. In a nutshell, the parametric PINN model is built of such fundamental operations whose derivative is well defined in terms of parameters themselves. Hence, one can compute the derivative of the whole PINN function by applying chain rule and using the derivatives of the basic operations.

\subsection{Measure of PINN accuracy - coefficient of determination}
\label{sec:r2_score}
The accuracy of PINN prediction is measured in terms of coefficient of determination, popularly known as R\textsuperscript{2} score. It is a measure of the amount of variance in the true target variable captured by the predicted values. Mathematically, it can be defined as:

\begin{equation}
    R^2 \text{ score} = 1 - \dfrac{\sum_{i\in \Omega} (u_i - \hat{u_i})^2}{\sum_{i\in \Omega} (u_i - \bar{u})^2}
    \label{eq:r2_score}
\end{equation}
\noindent where $\Omega$ represents the domain. $u_i$, $\bar{u}$ and $\hat{u_i}$ are scalar values and represent true displacement values, mean of true displacement values, and predicted displacement values. The coefficient of determination can similarly be obtained for any scalar variable of interest such as a particular strain component. While R\textsuperscript{2} score is in general captures the performance of PINN, it may not be a suitable measure in some cases. More disucssion on R\textsuperscript{2} score is provided in \Cref{sec:results_and_discussion}.

\section{Results and Discussion}
\label{sec:results_and_discussion}
\subsection{1D-FGM-ELAS-DIRCH}
\label{sec:1d_fgm_elas_dirch}
Consider a bar with uniform cross-section and unit length lying along the $x$-axis from $x=0$ to $x=1$. The bar is assumed to be functionally graded, i.e. the elastic modulus varies spatially as $E(x) = 1/(1+x)$. Moreover, the bar is assumed to be fixed at one end, and a load is applied on the other to observe the elastic response. Hence, for a displacement controlled analysis, the boundary conditions are given as $u(0) =0, u(1) = 1$, where $u(x)$ represents the displacement field of the bar as a function of $x$ under the given load. The differential equation governing the elastic response of object for given properties and boundary conditions is given in \Cref{tab:diff_loss_functions}. The solution $u({x})$ to the differential equation can be analytically derived to be:
\begin{equation}
    u(x) = \dfrac{x^2+2x}{3}
\end{equation}

\begin{figure}[h]
    \centering
    \includegraphics[width=0.7\textwidth]{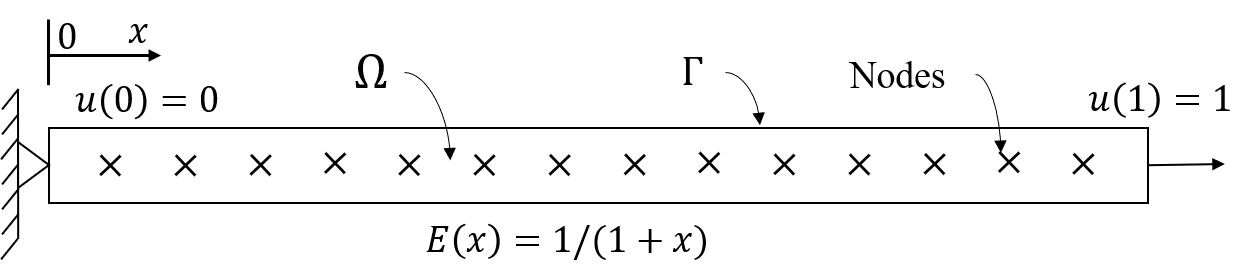}
    \caption{Schematic of one-dimensional bar with spatially varying stiffness and displacement node.}
    \label{fig:1d_schematic}
\end{figure}

\begin{table}[h]
\centering
\caption{Differential equations and loss functions used in training PINNs for solving different problems.}
\resizebox{\textwidth}{!}{
\begin{tabular}{|l|c|c|}
\hline
Problem Code & Differential Equation & Loss function  \\ \hline
1D-FGM-ELAS-DIRCH&$\dfrac{d}{d x}\left( \dfrac{1}{1 + x} \dfrac{d u}{d x}\right)=0$&  $\dfrac{1}{N}\sum_{i}^N\dfrac{1}{2(x_i+1)}\left(\left.\dfrac{d {u}}{dx}\right\rvert_{x_{i}}\right)^2$\\ \hline

1D-FGM-ELAS-NEU&$\dfrac{d}{d x}\left( \dfrac{1}{1 + x} \dfrac{d u}{d x}\right)=0$&$\dfrac{1}{N}\sum_{i}^N\dfrac{1}{2(x_i+1)}\left(\left.\dfrac{d{u}}{d x}\right\rvert_{x_{i}}\right)^2 - t(1){u}(1)$ \\ \hline 

1D-ELAS-BF&$\dfrac{d^2 u}{d^2 x} + 1= 0$  &$\dfrac{1}{N}\sum_{i}^N\left(\dfrac{1}{2}\left(\left.\dfrac{d {u}}{d x}\right\rvert_{x_{i}}\right)^2  - u_i\right)- t(1)u(1)$ \\ \hline 

1D-FGM-ELAS-THERMO&$\dfrac{d}{d x}\bkt{\dfrac{1}{1+x}\bkt{\dfrac{d u}{d x} - \alpha (T - T_0)} }= 0$&\multirow{2}{*}{$\dfrac{1}{N}\sum_{i=1}^N\dfrac{1}{2(1+x_i)}\bkt{\left.\dfrac{d u}{d x}\right|_{x_{i}} - T(x_i)}^2 + \dfrac{10}{(1+x_i)}\bkt{\left.\dfrac{d T}{d x}\right|_{x_{i}}}^2 $} \\
&$\dfrac{d }{d x}\bkt{k(x)\dfrac{\partial T}{\partial x}} = 0$& \\ \hline

Kirsch's Problem&$\nabla \cdot \mathbf{\sigma(x)} = 0$&$\sum_{\mathbf{x}_i \in \Omega}{\dfrac{1}{2}w(\mathbf{x}_i)\varepsilon(\mathbf{x}_i):\mathrm{C}:\varepsilon(\mathbf{x}_i)} - \sum_{\mathbf{x}_i \in \Omega_u} w'(\mathbf{x}_i)\mathbf{t}(\mathbf{x}_i)\cdot \mathbf{n}(\mathbf{x}_i) $ \\ \hline

2D-FGM-ELAS-NEU&$\nabla \cdot \mathbf{\sigma(x)} = 0$&$\sum_{\mathbf{x}_i \in \Omega}{\dfrac{1}{2}w(\mathbf{x}_i)\varepsilon(\mathbf{x}_i):\mathrm{C}(\mathbf{x}_i):\varepsilon(\mathbf{x}_i)} - \sum_{\mathbf{x}_i\in \Omega_{\Gamma_t}} \mathbf{t(x_i)}\cdot \mathbf{n(x_i)}$\\ \hline

2D-FGM-ELAS-DIRCH&$\nabla \cdot \mathbf{\sigma(x)} = 0$&$\sum_{\mathbf{x}_i \in \Omega}{\dfrac{1}{2}w(\mathbf{x}_i)\varepsilon(\mathbf{x}_i):\mathrm{C}(\mathbf{x}_i):\varepsilon(\mathbf{x}_i)}$ \\ \hline

2D-FGM-THERMO-ELAS&$\nabla \cdot \mathbf{\sigma(x)} =0$&\multirow{2}{*}{$\sum_{\mathbf{x}_i \in \Omega}\bkt{{\dfrac{1}{2}w(\mathbf{x}_i)\varepsilon^{\text{el}}(\mathbf{x}_i):\mathrm{C}(\mathbf{x}_i):\varepsilon^{\text{el}}(\mathbf{x}_i)} + \dfrac{1}{2}w(\mathbf{x}_i) \dfrac{10}{1+x}\nabla T(\mathbf{x}_i) \cdot \nabla T(\mathbf{x}_i)}$} \\
&$\nabla \cdot \bkt{k(\mathbf{x})\nabla T} = 0$& \\ \hline
\end{tabular}}
\label{tab:diff_loss_functions}
\end{table}
It is to be noted that throughout this study, consistent units of various physical quantities can be assumed. Hence, we do not explicitly mention units in various figures and tables. Now, to solve the above differential equation using a PINN, one needs to understand the three fundamental elements of the PINN framework: (1) training dataset, (2) choice of neural network architecture, (3) loss function. The training dataset or the input to the PINN will be the spatial coordinate of nodes in the domain. Hence, the domain has to be discretized into a set of nodes; however, unlike FEM, there is no notion of elements in the PINN framework. While there is the freedom to non-uniformly discretize the domain into nodes, for simplicity, the nodes are chosen to be uniformly distributed in this example. It is to be noted that there is no target dataset in training PINN models.
\begin{figure}[h]
    \centering
    \includegraphics[width=90mm]{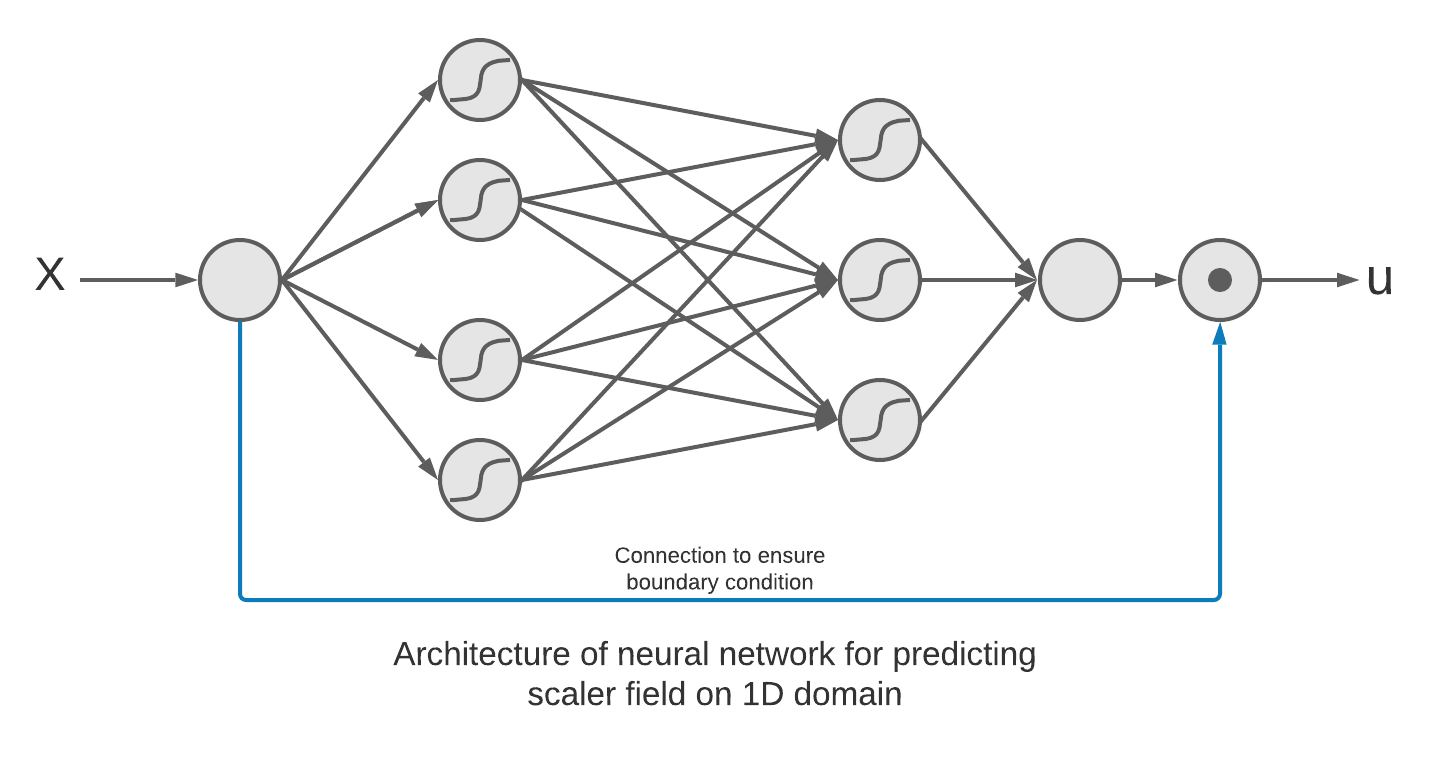}
    \caption{Representative PINN architecture for purely displacement controlled analysis of 1D material.}
    \label{fig:pinn_1d}
\end{figure}

\noindent The next step is to decide the neural network architecture. While the PINN framework offers a lot of flexibility in determining the model architecture, a few of such parameters get determined by the task at hand. For example, the number of nodes in the input and output layers of the model must respectively be equal to the dimensionality of the problem. In the case of the one-dimensional solid mechanics problem, both the input and output layer has one neuron each. The input neuron corresponds to the spatial position of nodes, and the output neuron corresponds to the displacement. Moreover, in the case of purely Dirichlet boundary conditions, the PINN model architecture can be modified to satisfy displacement boundary conditions implicitly (see~\Cref{eq:dirichlet}). 
    
 Let us assume that there is a blackbox neural network function $\tilde{v}(x):\mathrm{\Omega} \to \mathcal{R}$, as defined in \Cref{eq:nn}, which maps points in the spatial domain $\mathrm{\Omega}$ to a real number.
 Now, the Dirichlet boundary conditions as assumed in this example can be implemented by the transformation $\tilde{u}(x) := x + x(1-x)\tilde{v}(x)$ thereby ensuring $\tilde{u}(0) = 0 $ and $ \tilde{u}(1) = 1$. In \Cref{fig:pinn_1d}, connections in black colour correspond to $\tilde{u}(x)$, while the connection in blue is the transformation applied over $\tilde{v}(x)$ to get $\tilde{u}(x)$. The details of neural network $\tilde{v}(x)$ is given in \Cref{tab:architecture_details}. It is to be noted that one \textit{cannot} do any transformation to implement Dirichlet BC such as $\tilde{u}(x) := x^2 + x^2(1-x^2)\tilde{v}(x)$. This transformation will not only ensure $\tilde{u}(0) = 0$ and $\tilde{u}(1) = 1$, but it will also simultaneously put constraints on the displacement gradient which is $\tilde{u}'(0) = 0$ for the given case. Hence, a transformation which puts undesired constraints on gradients of primary variables must be avoided.

\begin{table}[h]
\centering
\caption{Details of the neural network architecture $(\mathbf{\tilde{v}}(\mathbf{x}))$ used in various problems.}
\resizebox{\textwidth}{!}{
\begin{tabular}{|l|c|c|c|c|c|}
\hline
 Problem Code & \begin{tabular}[c]{@{}c@{}}No. of\\ input nodes\end{tabular} & \begin{tabular}[c]{@{}c@{}}No. of\\ output nodes\end{tabular} & \begin{tabular}[c]{@{}c@{}}No. of\\ hidden layers\end{tabular} & \begin{tabular}[c]{@{}c@{}}No. of nodes\\ in hidden layers\end{tabular} & \begin{tabular}[c]{@{}c@{}}Activation\\ function\end{tabular} \\ \hline
1D-FGM-ELAS-DIRCH & 1 & 1 & 2 & [5, 5] & tanh(.) \\ \hline
1D-FGM-ELAS-NEU & 1 & 1 & 2 &  [5, 5] & tanh(.) \\ \hline
1D-ELAS-BF & 1 & 1 & 2 & [5, 5] & tanh(.) \\ \hline
1D-FGM-THERMO-ELAS & 1 & (1,1) & (2,1) &  [(5, 5), (10, 10), (5,5)]& tanh(.) \\ \hline
Kirch's Problem & 2 & 2 & 3 & [100, 100, 100] & tanh\textsuperscript{2}(.) \\ \hline
2D-FGM-ELAS-NEU & 2 & 2 & 3 & [100, 100, 100] & tanh\textsuperscript{2}(.) \\ \hline
2D-FGM-ELAS-DIRCH & 2 & 2 & 3 & [100, 100, 100] & tanh\textsuperscript{2}(.) \\ \hline
 2D-FGM-THERMO-ELAS & 2 & 3 & 2 & [300, 300] & elu\textsuperscript{2}(.) \\ \hline
\end{tabular}}
\label{tab:architecture_details}
\end{table}
 The last and the most crucial element of the PINN framework is the loss function used in training such models. As highlighted earlier, the loss used sets the PINN model apart from the rest of the neural network family. The PINN loss function is discretized form of the functional corresponding to the underlying differential equation. Hence, the loss function will also change from one differential equation to the other. It is to be noted that discretization is done to numerically evaluate the functional in \Cref{eq:functional} for training the PINN. The neural network model trains itself (fits the differential equation) by minimizing the loss function and thereby finding a function that minimizes the functional corresponding to the differential equation. However, the tricky part is that the functional and hence the loss function often depends not only on the primary target variables but also on their derivatives. Differential equation and corresponding loss function can be found in table \Cref{tab:diff_loss_functions}. 
  
\begin{figure}[h]
    \centering
    \includegraphics[width=\textwidth]{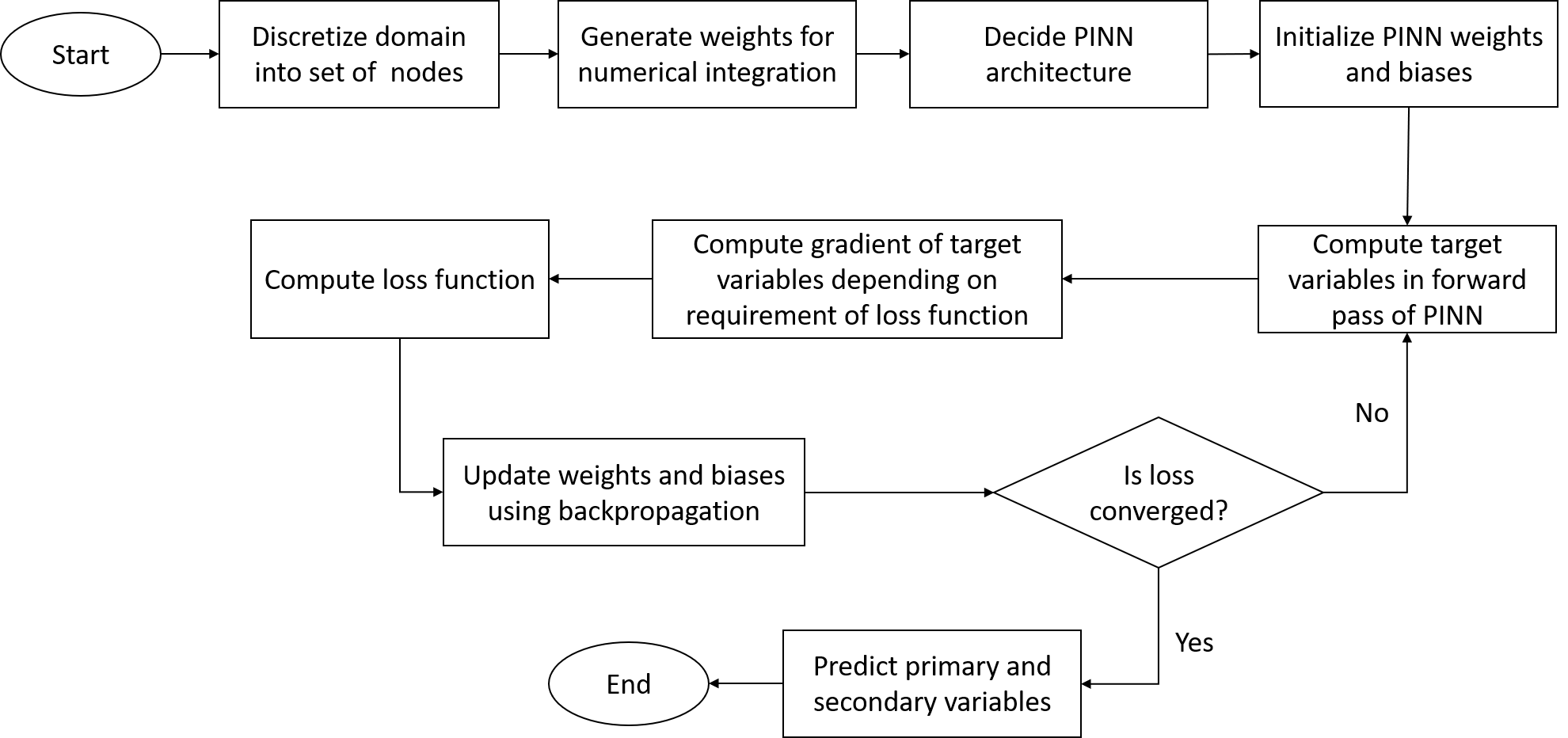}
    \caption{Flowchart of PINN framework}
    \label{fig:flowchart}
\end{figure}

\noindent In the context of solid mechanics, the loss function is the internal strain energy minus work done by external forces on the object under consideration. It is evident from \Cref{tab:diff_loss_functions}, the loss involves computing strain values which depends on gradient of $u(x)$ with respect to $x$. Since the neural network is a parametric composite function, it is possible to compute the derivative of the target variable $\tilde{u}(x)$ with respect to the input variable $x$ in terms of parameters of the neural network. While computing such derivatives can be a tedious task given the complexity of the neural network, the state-of-the-art deep learning libraries turn out to be the savior. The derivative can be easily computed using the automatic differentiation technique (discussed in~\Cref{sec:training_neural_network}) implemented using the library Tensorflow. The complete pipeline of the PINN framework is summarized in \Cref{fig:flowchart}.

\begin{figure}[h]
    \centering
    \includegraphics[width=140mm]{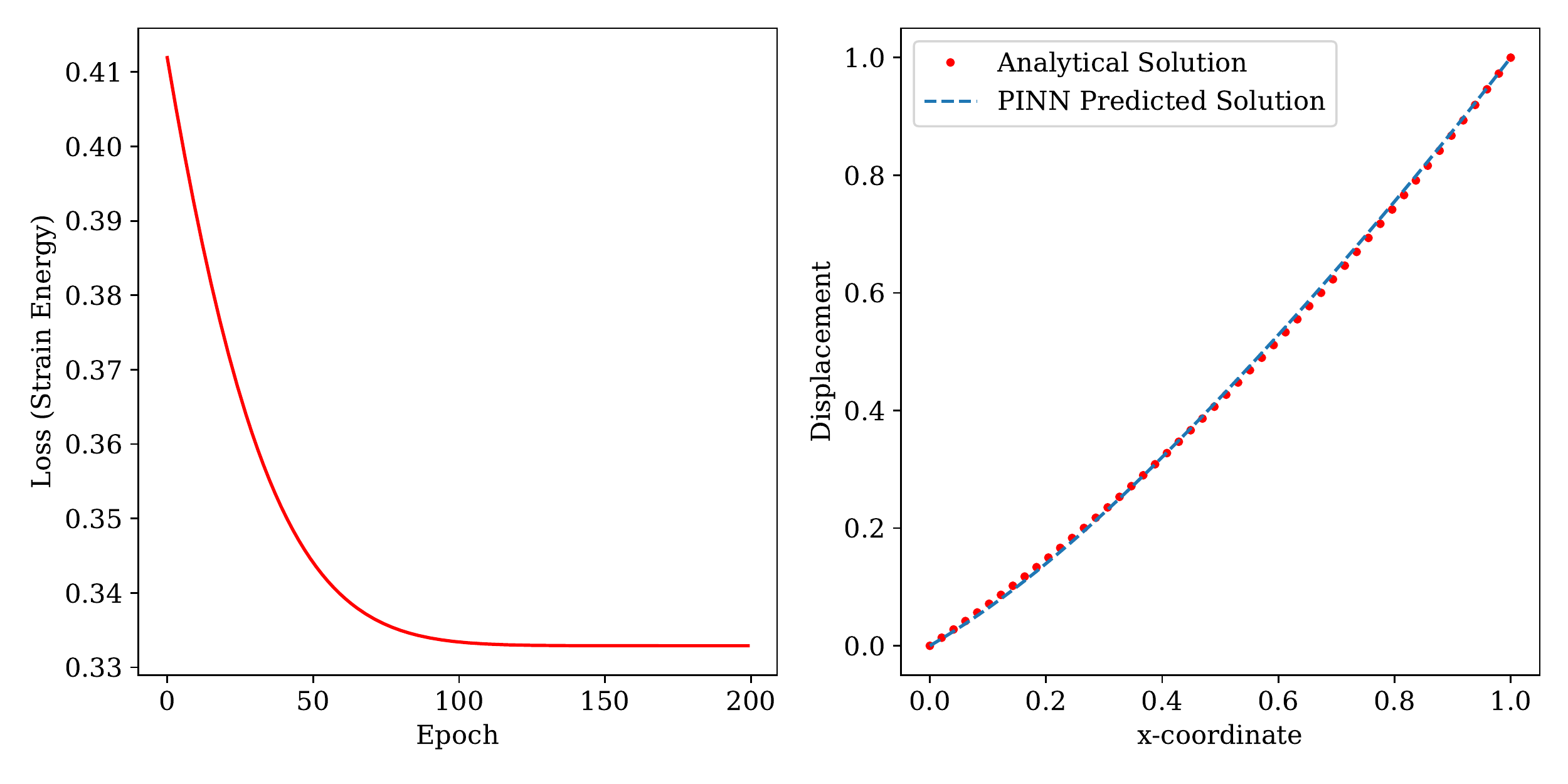}
    \caption{Decline in physics-informed loss (left) and comparison between analytical and predicted solution obtained at the end of training process (right) for one-dimensional elasticity problem with displacement controlled load (\Cref{sec:1d_fgm_elas_dirch}).}
    \label{fig:pinn_1d_dirchlet_fgm}
\end{figure}

\noindent After training the neural network with physics-informed loss, the given one-dimensional differential equation is solved to obtain $\hat{u}(x)$ which is the same as $\tilde{u}(x)$ at the end of training PINN. The comparison between $\hat{u}(x)$, and analytical solution is shown in the \Cref{fig:pinn_1d_dirchlet_fgm}. It is worthwhile to reiterate that the physics-informed loss must converge to the minimum strain energy. The training of neural networks is an optimization task where the global minima should ideally correspond to the unique minimum energy solution of the system. R\textsuperscript{2} for different variables in given in \Cref{tab:r2_scores}.

\begin{table}[h]
\centering
\caption{Coefficient of determination (R\textsuperscript{2} score) for various variables of interest in different solid mechanics problems.}
\resizebox{\textwidth}{!}{
\begin{tabular}{|l|c|c|c|c|c|c|c|c|c|}
\hline
Problem Code & $u_1$ & $u_2$ & $\sigma_{11}$ & $\sigma_{12}$ & $\sigma_{22}$ & $\varepsilon^{el}_{11}$ & $\varepsilon^{el}_{12}$ & $\varepsilon^{el}_{22}$ &$T$\\ \hline
 1D-FGM-ELAS-DIRCH&99.94\%& - & $<0$ &-  &-  &95.14\%  &-  &- &- \\ \hline
 1D-FGM-ELAS-NEU&99.99\% & - & $<0$ & - & - &99.99\%  &-  &-&-  \\ \hline
  1D-ELAS-BF&99.98\%& - & 99.65\%  &-  &-  &99.65\%  &-  &- &- \\ \hline
 1D-ELAS-THERMO-ELAS& 99.99\% & - & $<0$ & - & - & 97.61\% & - & - &99.94\% \\ \hline
  Kirsch's Problem&99.89\%&99.02\%&97.52\%&99.29\%&89.75\%& 97.32\% & 99.29\% & 88.05\% & -\\ \hline
 2D-FGM-ELAS-NEU& 99.83\% & 99.99\% & 83.01\% & 80\% & $<0$ &97.11\% &64.11\% &96.85\% &- \\ \hline
 2D-FGM-ELAS-DIRCH& 99.92\% &99.99\%  &81.78\%  &90.81\%  &$<0$& 98.16\% & 87.34\% & 96.75\% & -\\ \hline
  2D-FGM-THERMO-ELAS& 99.99\% & 99.99\% & 88.84\% & 70.91\% & 73.92\% & 99.93\% & 99.95\% &99.95\% &99.97\% \\ \hline
\end{tabular}}
\label{tab:r2_scores}
\end{table}

\subsection{1D-FGM-ELAS-NEU}
\label{sec:1d_fgm_elas_neu}
The domain is again assumed to be spanning from $x=0$ to $x=1$, while the cross-section is uniform. With the same functionally graded elastic modulus, the differential equation governing the elastic response remains the same as the last problem (see~\Cref{tab:diff_loss_functions}). While the left end of the bar is fixed, a unit traction $t$ is applied on the right end such that $u(0) = 0, \quad t(1) = 1$. The analytical solution for this case is given as:
\begin{equation}
    u(x) = \dfrac{2x + x^2}{2}
\end{equation} 

The training dataset remains the same, i.e., the set of nodes on the domain. While the neural network architecture implicitly ensured the Dirichlet boundary conditions, implementation of Neumann boundary condition is ensured in the definition of loss function given in \Cref{tab:diff_loss_functions}. Since the loss function is internal energy minus the work done by external forces, unlike the other use cases of a neural network, the loss function does have a physical significance in the case of PINN. The loss can be expected to converge to zero only in situations where the internal energy stored is equal to the work done. Otherwise, it can very well either saturate to a negative value if work done by the external force is more than the energy stored or can saturate to a positive value if the loss is only composed of internal energy. It is also to be noted that loss function is not always 
internal energy minus the work done by external force even when the later can be a non-zero quantity. For example, in previous example the loss function was only composed of internal energy even though work done by external force would be finite. So whenever, the Dirichlet boundary condition is satisfied by a transformation abstracted in \Cref{eq:dirichlet}, the work done by the corresponding reaction force must not be subtracted in the loss function.

\begin{figure*}[t]
    \centering
    \includegraphics[width=140mm]{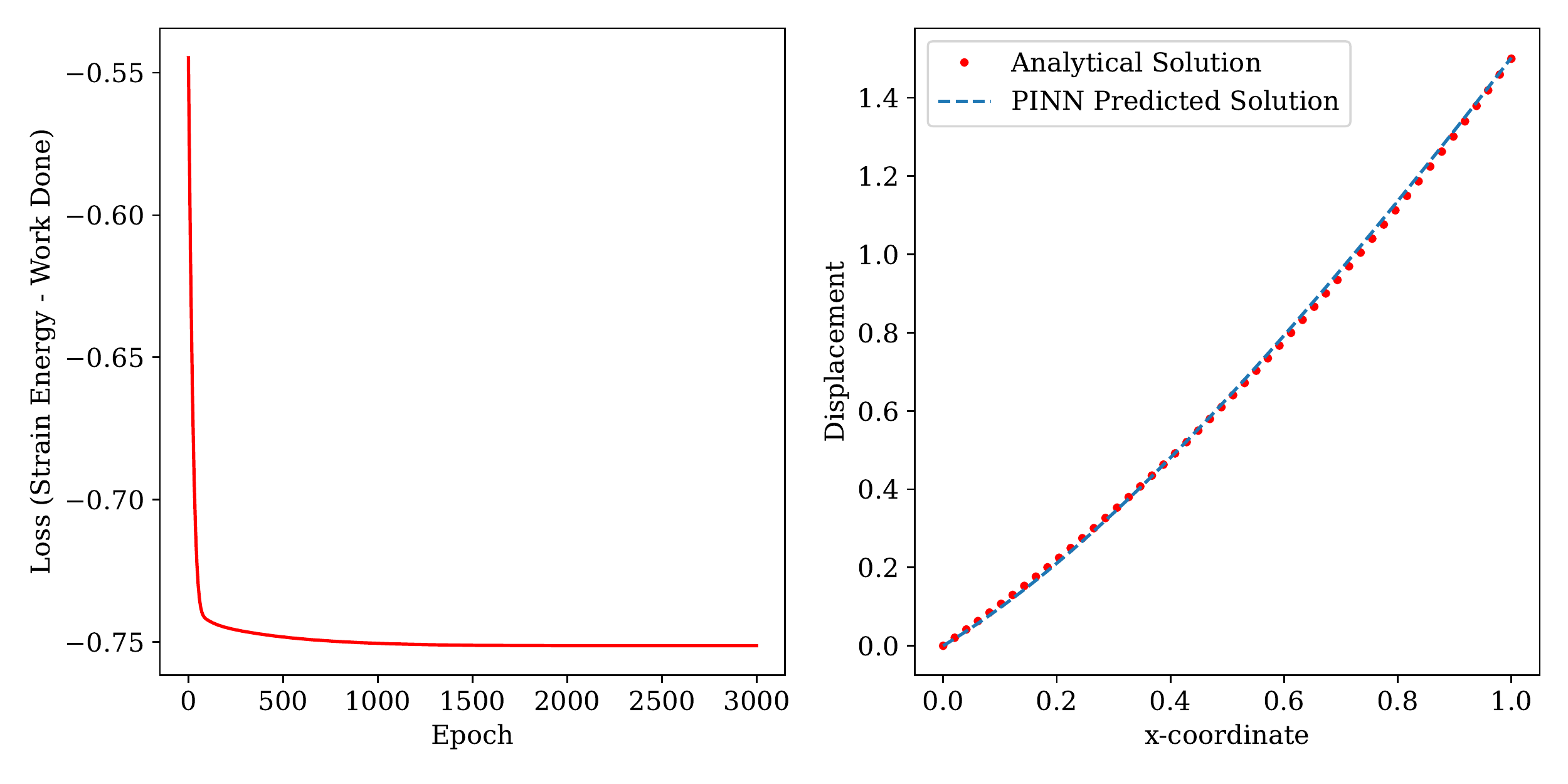}
    \caption{Decline in physics-informed loss (left) and comparison between analytical and predicted solution obtained at the end of training process (right) for one-dimensional elasticity problem with force controlled load (\Cref{sec:1d_fgm_elas_neu}).}
    \label{fig:pinn_1d_neumann_fgm}
\end{figure*}

The comparison between analytical and PINN solutions for this case is shown in \Cref{fig:pinn_1d_neumann_fgm}. It can be observed that the loss converges to a negative value as the work done by an external force is more than that of strain energy stored in the bar. Moreover, a good match is observed between the analytical and PINN predicted solution. The R\textsuperscript{2} score for different variables is given in Table~\ref{tab:r2_scores}.

\subsection{1D-ELAS-BF}
\label{sec:1d_elas_bf}
In the last two examples, the use of PINN to implement Dirichlet and Neumann boundary conditions was discussed. We consider one more example with body force to demonstrate the robustness of the PINN framework. In this case, we will assume the elastic modulus to be unity over the
domain. Other properties such as area of cross-section, mass density are assumed to be unity. The governing equations are given in \Cref{tab:diff_loss_functions}. Boundary conditions are applied such that $u(0) = 0, \quad t(1) = 1$. The analytical solution for this case is given by:
\begin{equation}
    u(x) = \dfrac{4x - x^2}{2}
\end{equation}

\begin{figure*}[h]
    \centering
    \includegraphics[width=140mm]{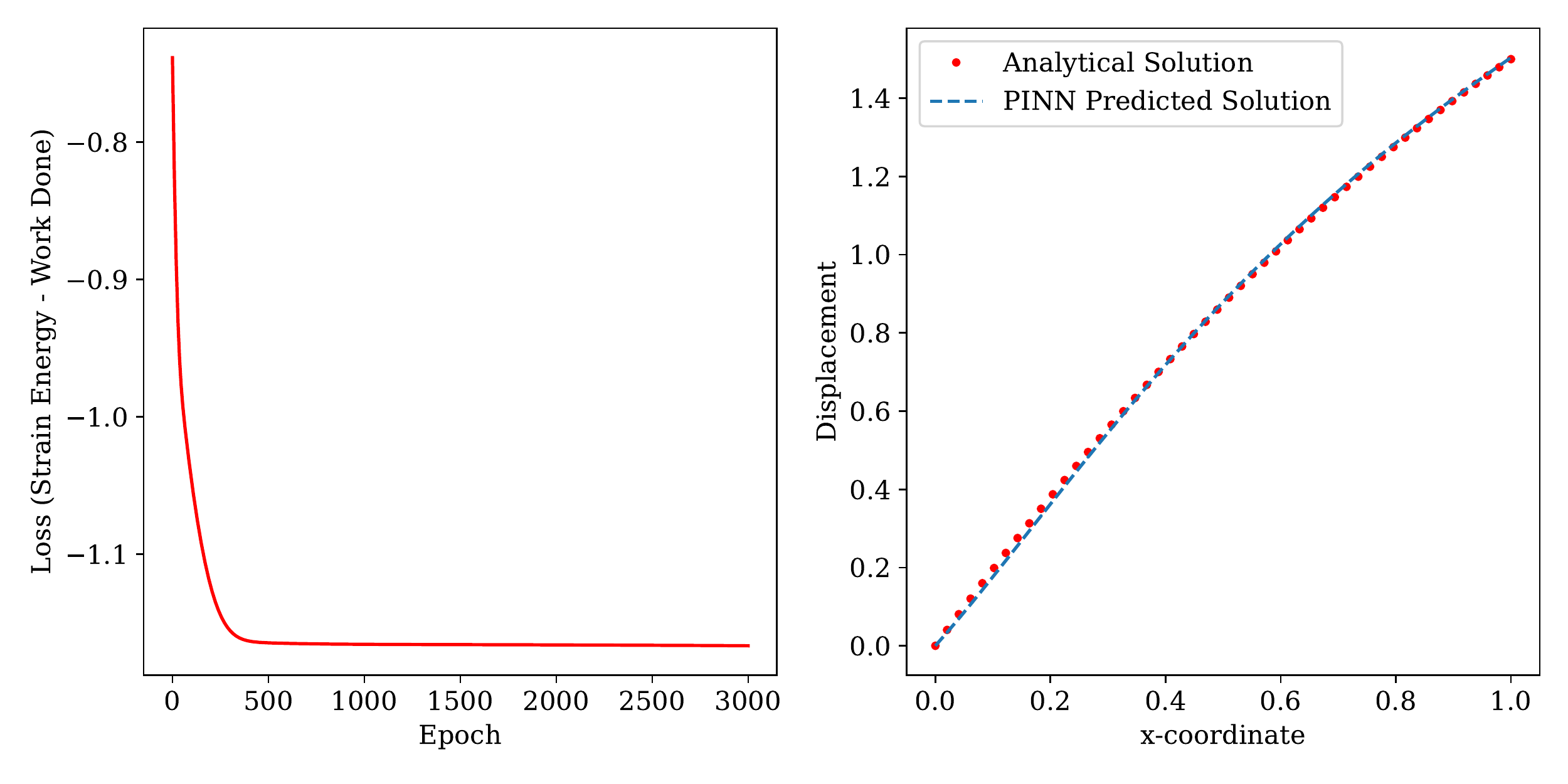}
    \caption{Decline in physics-informed loss (left) and comparison between analytical and predicted solution obtained at the training process (right) for 1D mechanics problem with force controlled analysis and body force (\Cref{sec:1d_elas_bf}).}
    \label{fig:pinn_1d_neumann_bf}
\end{figure*}

\noindent Unsurprisingly, the PINN model can match the analytical solution in this case as well. The comparison is shown in \Cref{fig:pinn_1d_neumann_bf} and R\textsuperscript{2} score was given in Table~\ref{tab:r2_scores}.

It is quite remarkable that the PINN model can perfectly fit solutions to differential equations under different boundary conditions. However, one cannot just expect the black-box PINN model to fit the solution of all possible differential equations. For example, a complex PINN model that is a non-linear function can never fit a differential equation where the solution $u$ is a constant value or linear in $x$. A practical example would be estimating the displacement field in a coarsely discretized multi-phase composite microstructure. Due to the coarse discretization, the true displacement would be piece-wise linear functions. Moreover, the gradient at the interface of two adjacent phases can be very high depending on the difference in their stiffness. Hence, it will be difficult for PINN to fit that displacement field with a high displacement gradient at the interface of phases. However, this inability of PINN need not necessarily be a disadvantage because coarse discretization is only a problem pertaining to the modeling of materials. In reality, displacement fields in materials will be reasonably smooth unless there is a discontinuity in the material due to cracks or manufacturing defects.

It is to be noted that all the discussed examples were selected such that the solution is quadratic in nature. While it is often challenging to comment on the nature of the black-box PINN function, it can undoubtedly be said from the excellent match that the models were just aptly parameterized to fit solutions of quadratic nature.
\subsection{1D-FGM-THERMO-ELAS}
\label{sec:1d_fgm_thermo_elas}
\noindent In this section, we demonstrate the ability of PINN to estimate thermo-mechanical response of the functionally graded material. Let's again consider a 1D bar $\mathbf{\Omega}$ from $x=0$ to $x=1$. The thermo-elastic response of the object is governed by the differetial equations given in \Cref{tab:diff_loss_functions}. Let the elastic modulus of object vary as $E(x) = {1}/\bkt{1+x}$ and conductivity to vary as $k(x) = {10}/\bkt{1+x}$. The mechanical and thermal boundary conditions are imposed such that $u(0) = 0;\quad u(1) = 1;\quad T(0) = 0;\quad T(1) = 1$ where $u(x)$ and $T(x)$ are respectively the displacement and temperature fields. The coefficient of thermal expansion $\alpha$ and reference temperature $T_0$ are assumed to be unity and zero respectively. The analytical solution for the purpose of comparison can be derived to be:
\begin{equation}
    u(x) = \dfrac{x^3}{9} + \dfrac{14}{27}x^2 + \dfrac{10}{27}x
\end{equation}
and
\begin{equation}
    T(x) = \dfrac{2x+x^2}{3}
\end{equation}

\begin{figure}[h]
    \centering
    \includegraphics[width=140mm]{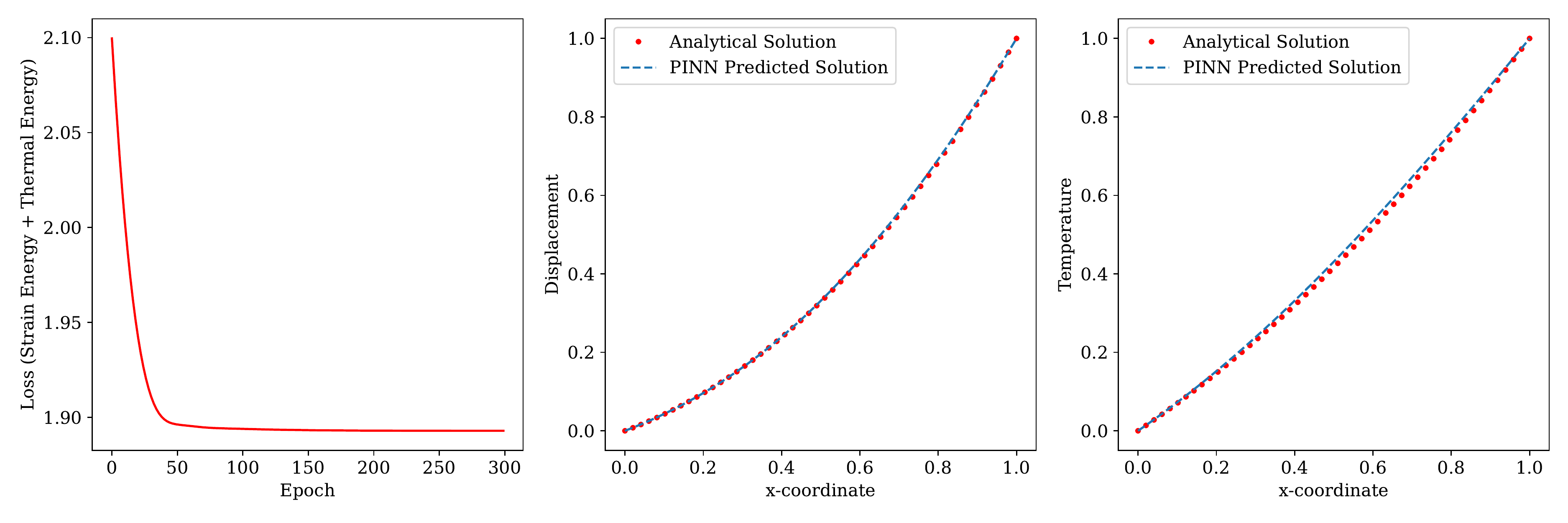}
    \caption{Reduction is loss with epoch (left), comparison of analytical solutions with predicted displacement field (middle) and temperature field (right) obtained at the end of training process (\Cref{sec:1d_fgm_thermo_elas}). }
    \label{fig:thermo_mech_1d_dirichlet}
\end{figure}

The usual steps described in last section are followed in order to solve the pair of differential equations using PINN. The domain is discretized into a set of uniformly spaced $N=50$ nodes. The neural network $\tilde{\mathbf{v}}(x) = \bkt{ \tilde{v_1}(x)\quad \tilde{v_2}(x)}^T$, is transformed as follows to ensure satisfaction of Dirichlet boundary conditions:
\begin{equation}
    \begin{bmatrix}
    \tilde{u}(x) \\ \tilde{T}(x) 
    \end{bmatrix}
    = 
    \begin{bmatrix}
    x + (1-x)\tilde{v}_1(x) \\
    x + (1-x)\tilde{v}_2(x) \\
    \end{bmatrix}
\end{equation}
  The neural network architecture of $\mathbf{v}(x)$ is given in \Cref{tab:architecture_details}. Further, the loss for training the PINN is given in \Cref{tab:diff_loss_functions}. The model is sufficiently trained and comparison between analytical and PINN predicted solutions can be observed in \Cref{fig:thermo_mech_1d_dirichlet}. The R\textsuperscript{2} score for primary and secondary variables are given in \Cref{tab:r2_scores}. 

\subsection{Kirsch's Problem}
\label{sec:kirschs_problem}
\begin{figure}[h]
    \centering
    \includegraphics[width=90mm]{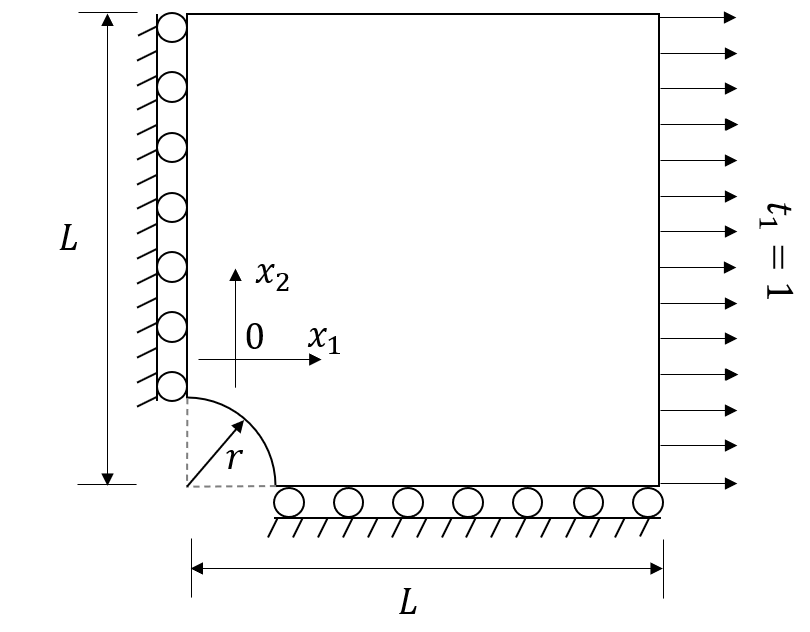}
    \caption{Domain, boundary conditions and applied load for Kirsch's problem (\Cref{sec:kirschs_problem}).}
    \label{fig:pwh}
\end{figure}

Let us consider an infinite rectangular plate with a circular hole. Let us assume that the circular hole has a radius of $0.1$ and is centered at $(0, 0)$. Moreover, the plate's thickness and the material's elastic modulus are assumed to be unity with appropriate units. Uniaxial traction $\mathbf{t}=[1 \quad 0]^T$ is applied in x-direction on both ends of the plate. The problem can be solved by considering a quarter of the infinite plate and applying Dirichlet boundary conditions as per symmetry. Moreover, for the purpose of simulation, the domain is assumed to be finite, as shown in \Cref{fig:pwh}.

\begin{figure}[h]
    \centering
    \includegraphics[width=90mm]{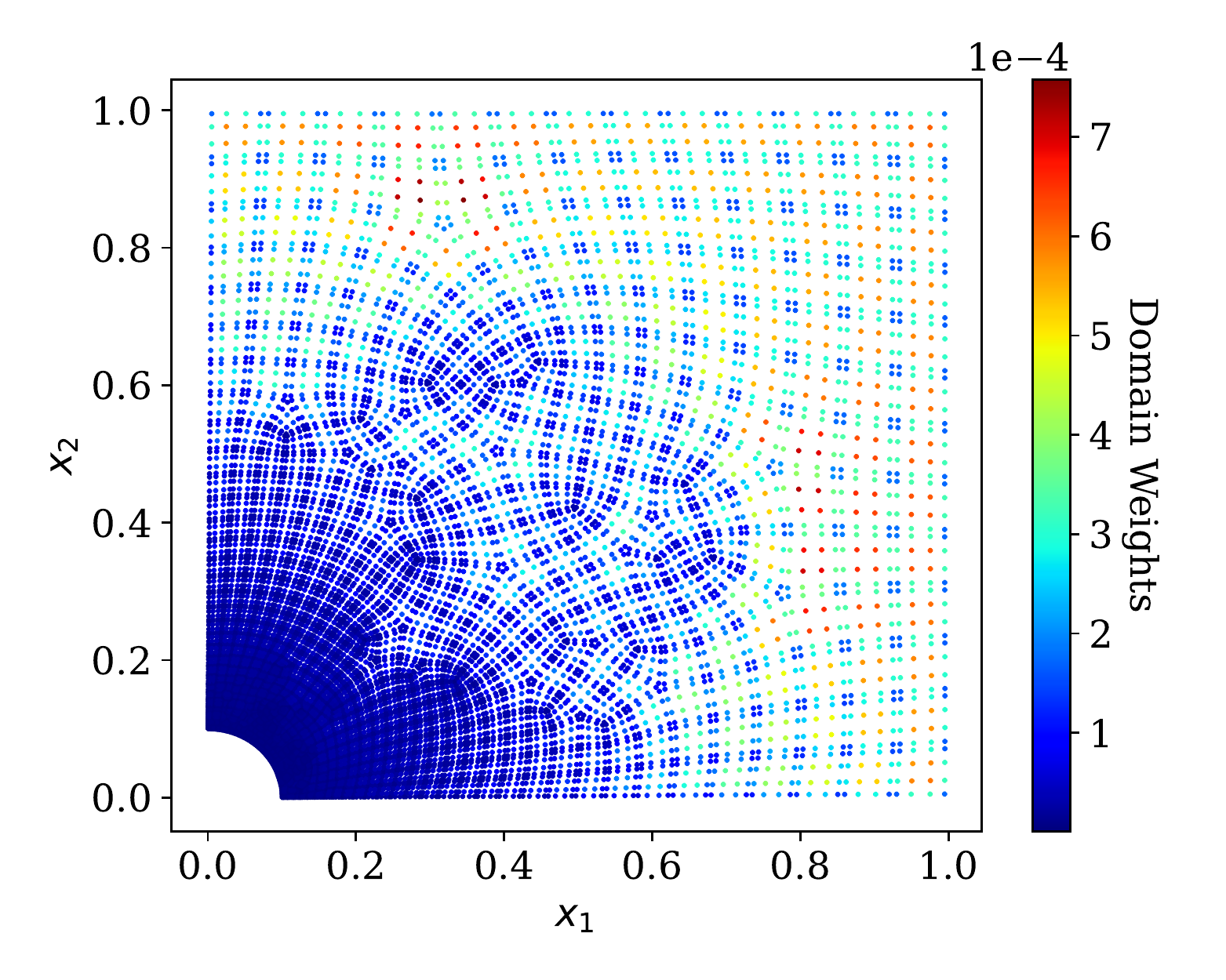}
    \caption{Distribution of nodes and weights in Kirsch's Problem (\Cref{sec:kirschs_problem}). Non-uniform distribution of nodes is necessary to model curvature of the hole.}
    \label{fig:pwh_mesh}
\end{figure}

Let us say that $\mathbf{\tilde{v}} (\mathbf{x}):\mathrm{\Omega} \to \mathcal{R}^2$ is the neural network model which maps 
spatial coordinates vector $(\mathbf{x})$ to a real-valued vector in $\mathcal{R}^2$. $\Omega$ represents the set of points that constitute the plate while $\Gamma$ is the set of points on the boundary of the plate. The details of neural network $\mathbf{\tilde{v}} (\mathbf{x}):\mathrm{\Omega} \to \mathcal{R}^2$ is given in Table~\ref{tab:architecture_details}.
To ensure the satisfaction of Dirichlet boundary conditions, the following transformation is applied $\mathbf{\tilde{u}}(\mathbf{x}) = \mathbf{x} \odot \tilde{v}(\mathbf{x})$, where $\mathbf{\tilde{u}({x})}$ gives displacement vector and $\odot$ represents Hadamard product. The governing equations and corresponding loss function is given in \Cref{tab:diff_loss_functions}. Under the plane stress conditions, the material constitutive matrix is given in \Cref{eq:stiffness_matrix}. Distribution of nodes and corresponding weights ($w(\mathbf{x}))$ are shown in the \Cref{fig:pwh_mesh}. It is to be noted that weights are required for numerical integration in case nodes are non-uniformly distributed over the domain. In the case of uniform distribution of nodes, as in the case of other examples, the weights would be a constant value.

\begin{figure}[h]
    \centering
    \includegraphics[width=90mm]{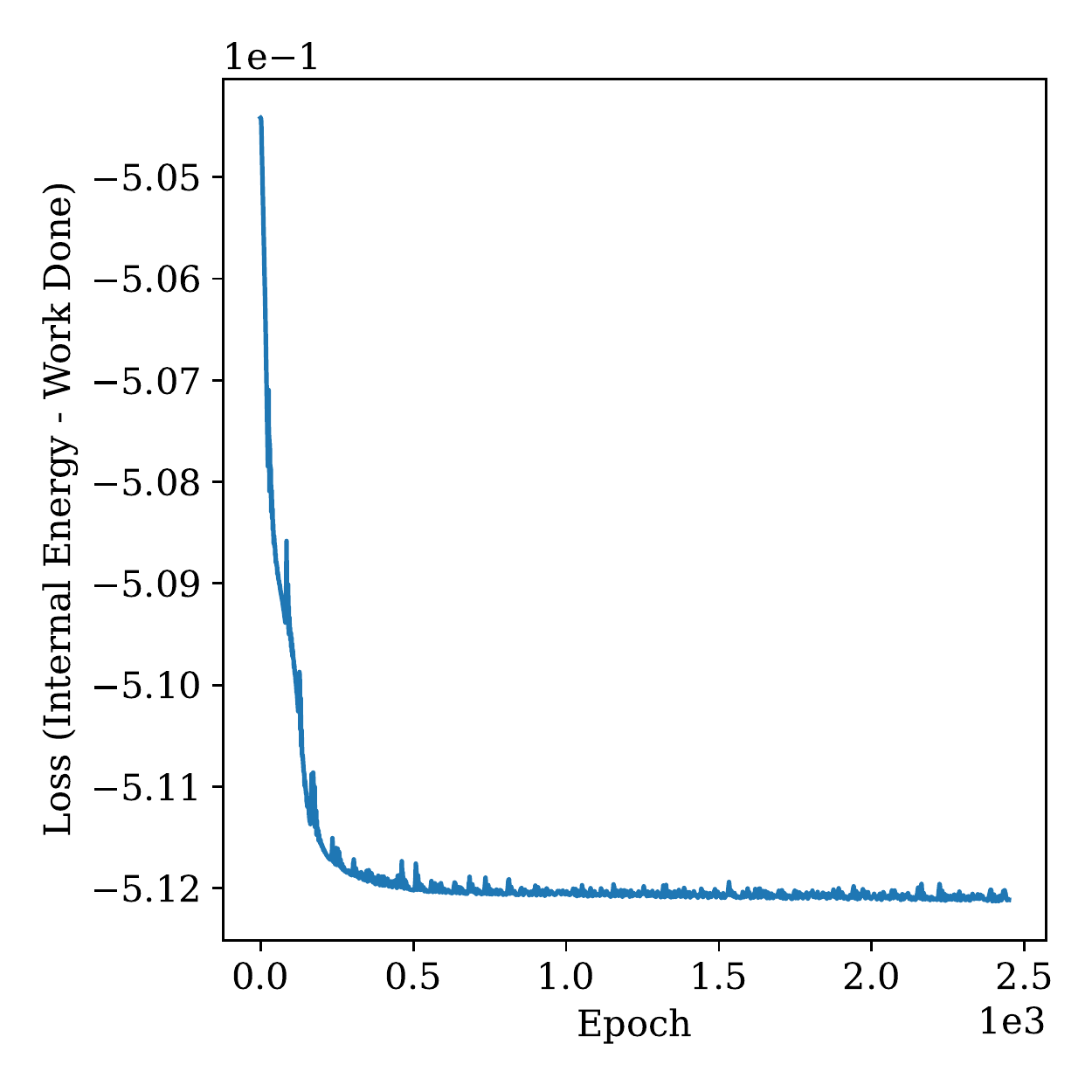}
    \caption{Reduction of loss vs epoch as the PINN trained in Kirsch's Problem (\Cref{sec:kirschs_problem}).}
    \label{fig:pwh_loss}
\end{figure}

\begin{figure}[h]
     \centering
     \begin{subfigure}[b]{\textwidth}
         \centering
         \includegraphics[width=0.7\textwidth]{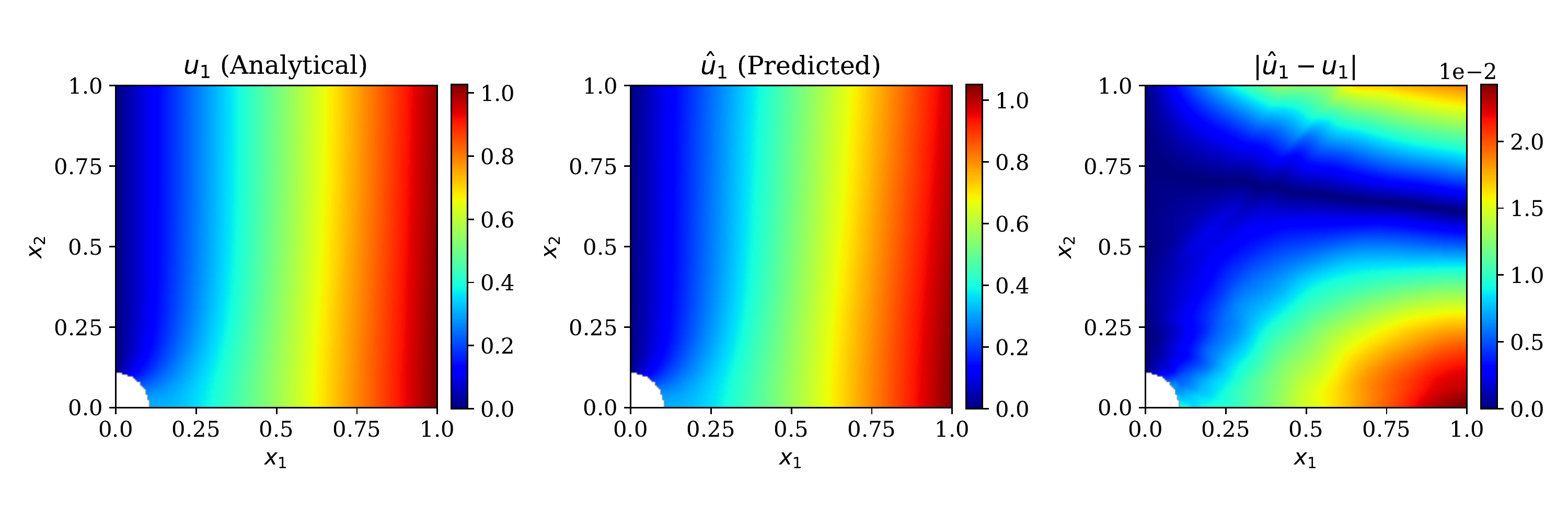}
         \caption{$u_1(\mathbf{x})$}
     \end{subfigure}
     \hfill
     \begin{subfigure}[b]{\textwidth}
         \centering
         \includegraphics[width=0.7\textwidth]{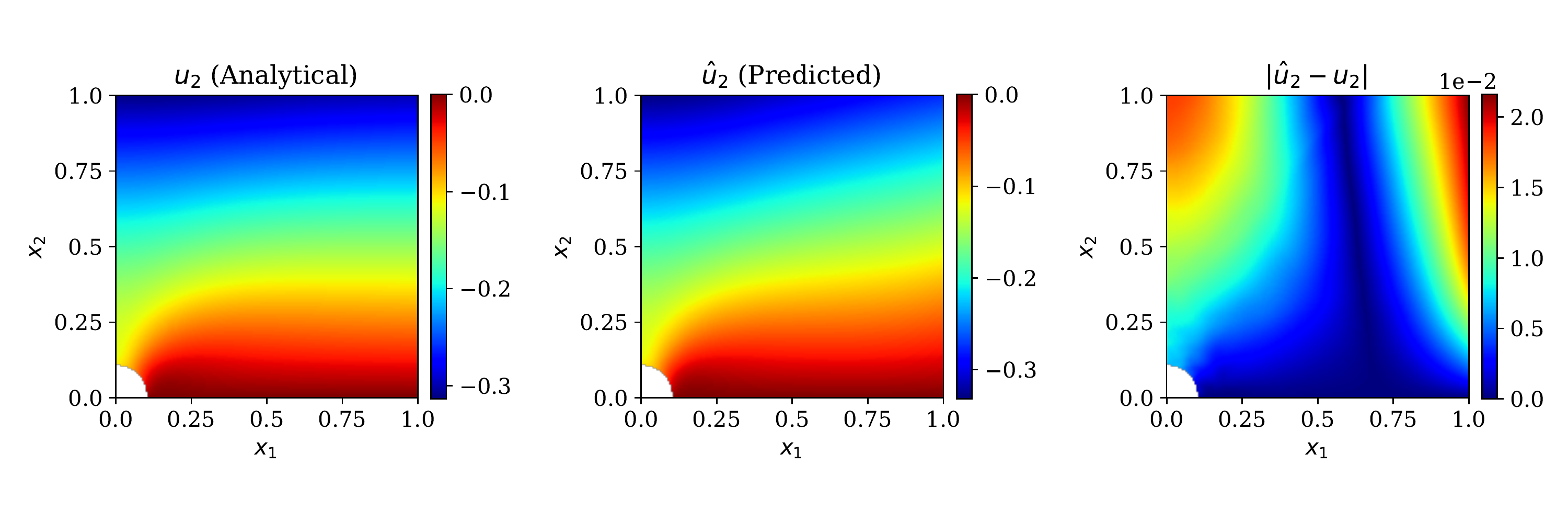}
         \caption{$u_2(\mathbf{x})$}
     \end{subfigure}
     \caption{Comparison between PINN predicted displacement field and corresponding analytical solution (\Cref{sec:kirschs_problem}).}
     \label{fig:kirsch_disp_comp}
\end{figure}

\begin{figure}[h]
     \centering
     \begin{subfigure}[b]{\textwidth}
         \centering
         \includegraphics[width=0.7\textwidth]{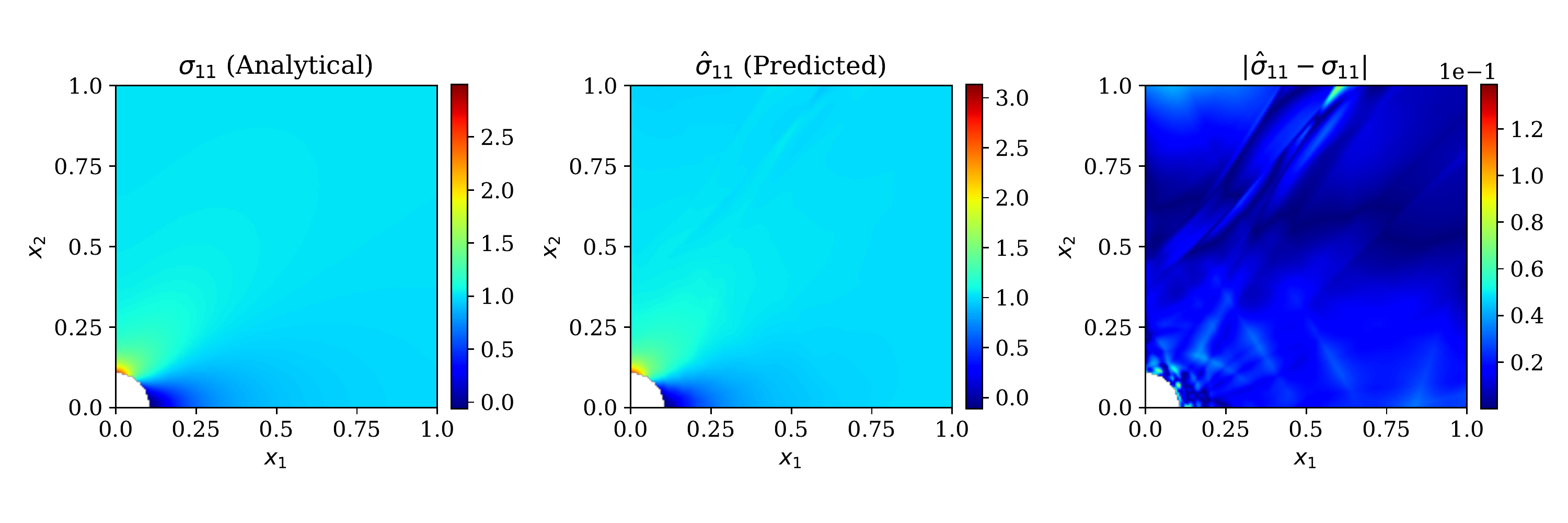}
         \caption{$\sigma_{11}(\mathbf{x})$}
     \end{subfigure}
     \hfill
     \begin{subfigure}[b]{\textwidth}
         \centering
         \includegraphics[width=0.7\textwidth]{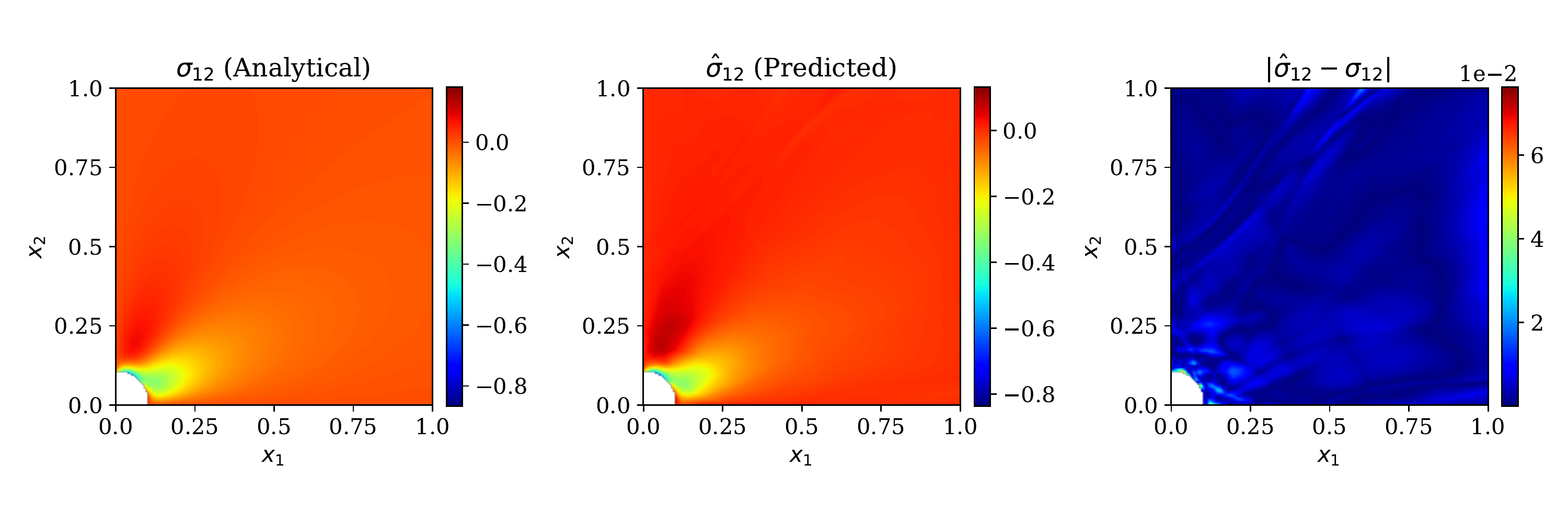}
         \caption{$\sigma_{12}(\mathbf{x})$}
     \end{subfigure}
     \hfill
     \begin{subfigure}[b]{\textwidth}
         \centering
         \includegraphics[width=0.7\textwidth]{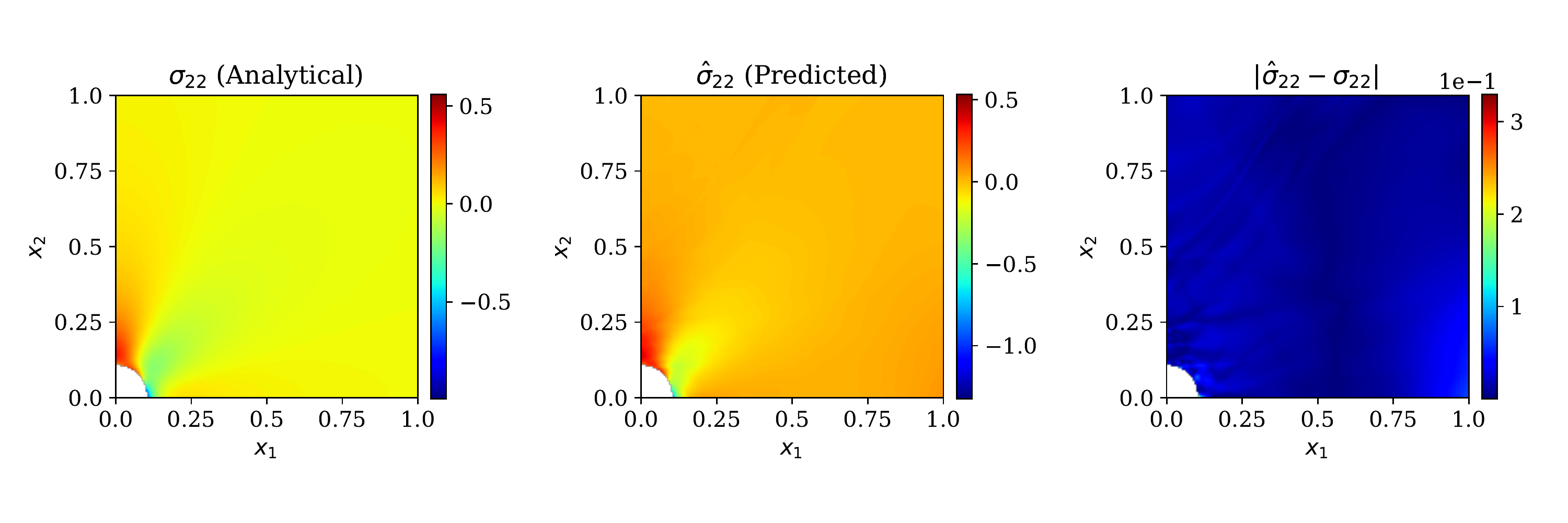}
         \caption{$\sigma_{22}(\mathbf{x})$}
     \end{subfigure}
        \caption{Comparison between PINN predicted stress fields and corresponding analytical solutions (\Cref{sec:kirschs_problem}).}
        \label{fig:kirsch_stress_comp}
\end{figure}

The PINN model is trained for 2500 epochs in this case. The loss, which is the internal strain energy - work done decreased rapidly at the beginning of the training process and then saturated to a value of approximately $-0.512$. The reduction in the loss vs. epochs is shown in \Cref{fig:pwh_loss}. In order to better understand the training process, it is worthwhile to look at the distribution of activation functions, weights and biases, and gradients for weight updates throughout the training process. They are given in Appendix: \Cref{fig:activation_distribution,fig:weight_distribution,fig:gradient_distribution}. The comparison between predicted and analytical solutions is shown in \Cref{fig:kirsch_disp_comp}, while the same for stress fields can be found in \Cref{fig:kirsch_stress_comp}. It is observed that the predicted fields are in good agreement with the analytical solution. The R\textsuperscript{2} score for various variables can found in \Cref{tab:r2_scores}.    

\subsection{2D-FGM-ELAS-NEU}
\label{sec:2d_fgm_elas_neu}
\begin{figure}[h]
    \centering
    \includegraphics[width=80mm]{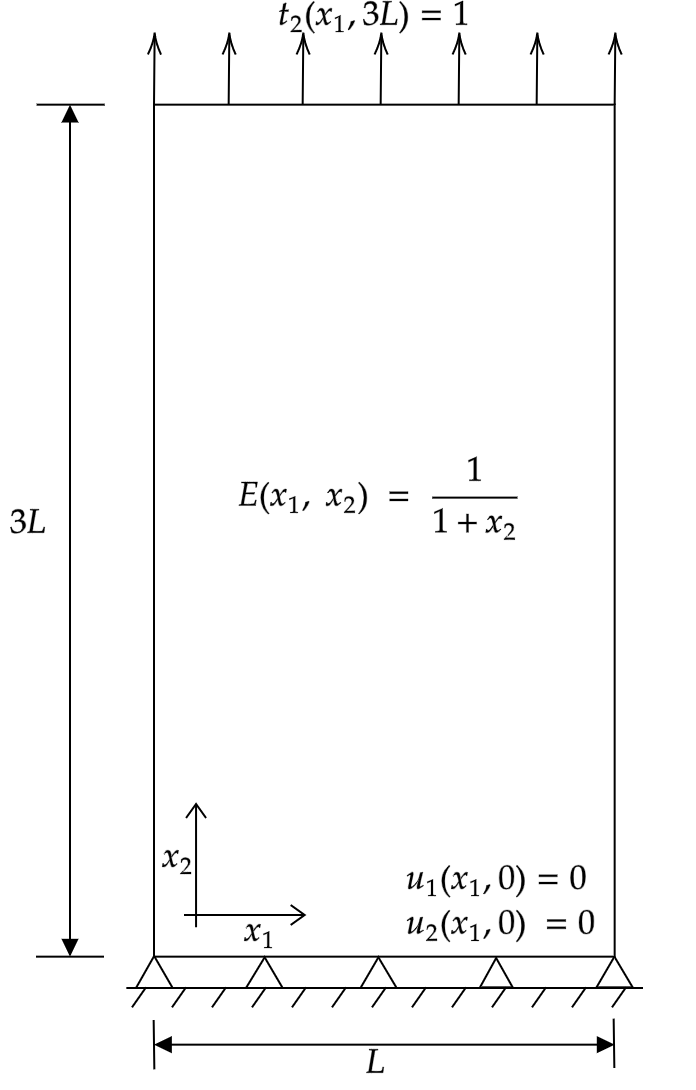}
    \caption{Domain and boundary conditions for functionally graded material object under force load defined in \Cref{sec:2d_fgm_elas_neu}.}
    \label{fig:neumann_fgm_2d}
\end{figure}

Let us consider a rectangular domain $\Omega$ as shown in the \Cref{fig:neumann_fgm_2d}. As shown in the figure, both the degrees of freedom  of nodes at the bottom edge are constrained to move. The elastic modulus of the plate is varying along $x_2$ direction and is given as $ E(x_1,x_2) = {1}/\bkt{1+x_2}$.

\begin{figure}[h]
    \centering
    \includegraphics[width=80mm]{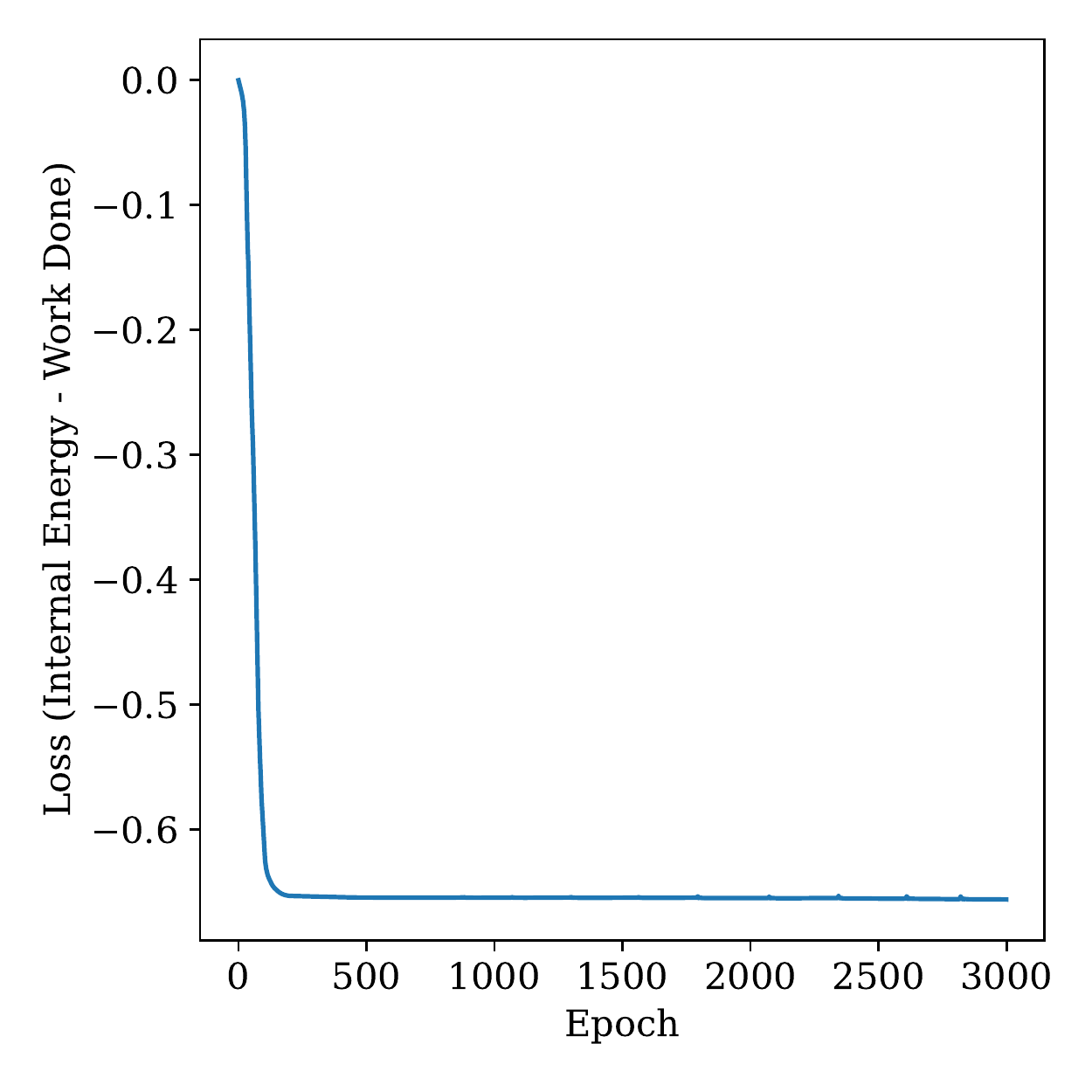}
    \caption{Reduction in loss as the PINN is trained to predict elastic response of two-dimensional functionally graded material object defined in \Cref{sec:2d_fgm_elas_neu}.}
    \label{fig:neumann_fgm_loss}
\end{figure}

\begin{figure}[h]
     \centering
     \begin{subfigure}[b]{\textwidth}
         \centering
         \includegraphics[width=0.7\textwidth]{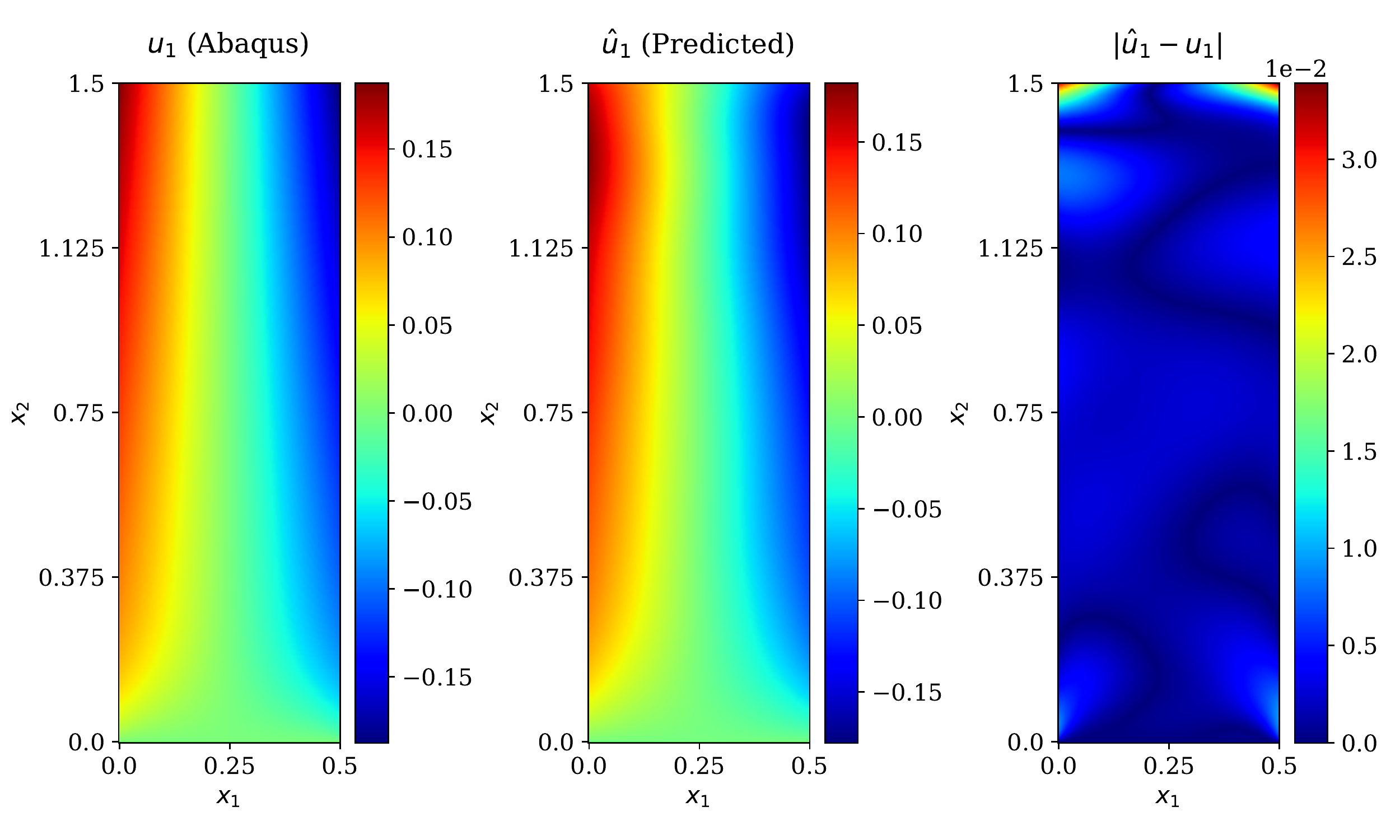}
         \caption{$u_1\bkt{\mathbf{x}}$}
     \end{subfigure}
     \hfill
     \begin{subfigure}[b]{\textwidth}
         \centering
         \includegraphics[width=0.7\textwidth]{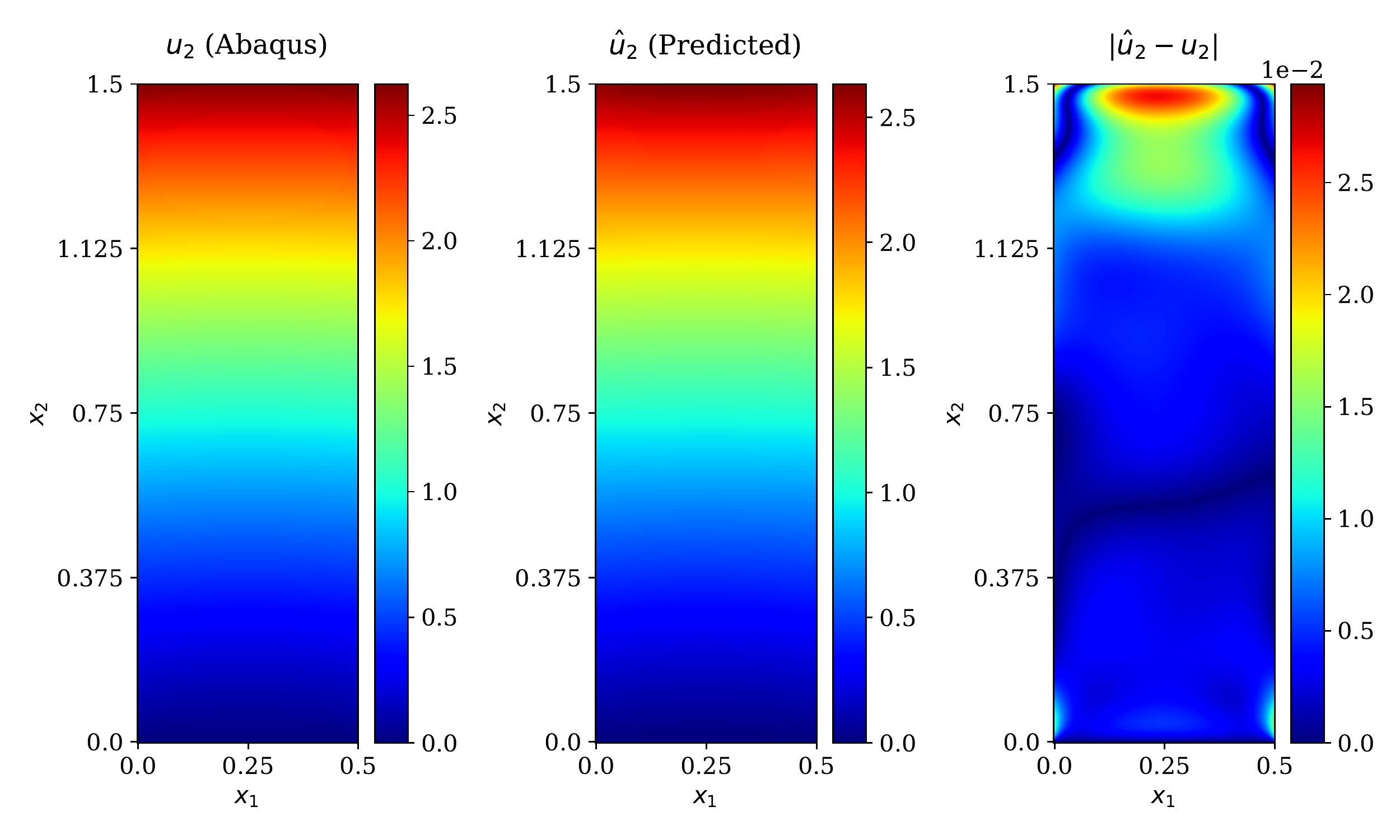}
         \caption{$u_2\bkt{\mathbf{x}}$}
     \end{subfigure}
     \caption{Comparison of PINN predicted displacement fields and corresponding Abaqus solution for problem defined in \Cref{sec:2d_fgm_elas_neu}.}
     \label{fig:2d_fgm_elas_neu_disp_comp}
\end{figure}

\begin{figure}[h]
     \centering
     \begin{subfigure}[b]{\textwidth}
         \centering
         \includegraphics[width=0.7\textwidth]{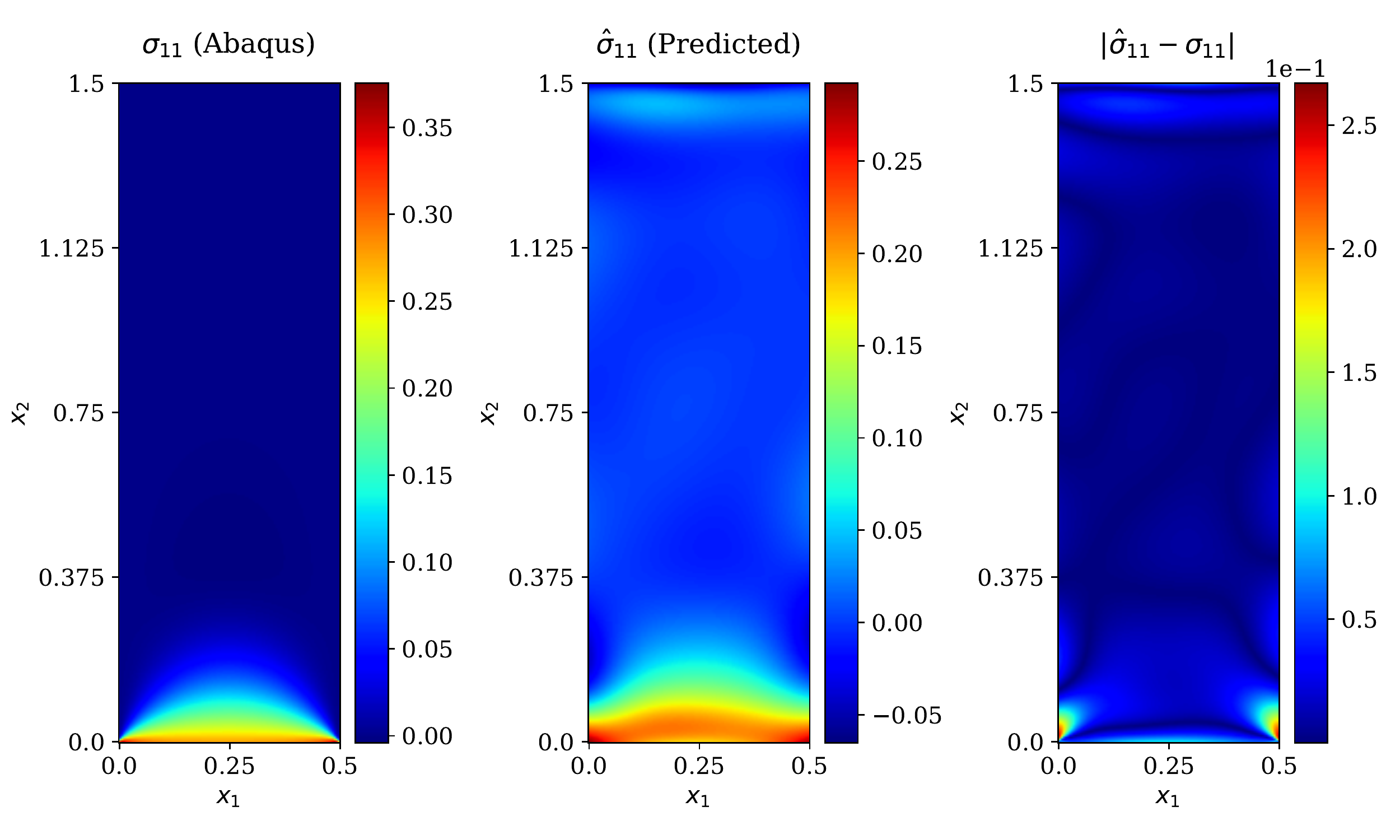}
         \caption{$\sigma_{11}\bkt{\mathbf{x}}$}
     \end{subfigure}
     \hfill
     \begin{subfigure}[b]{\textwidth}
         \centering
         \includegraphics[width=0.7\textwidth]{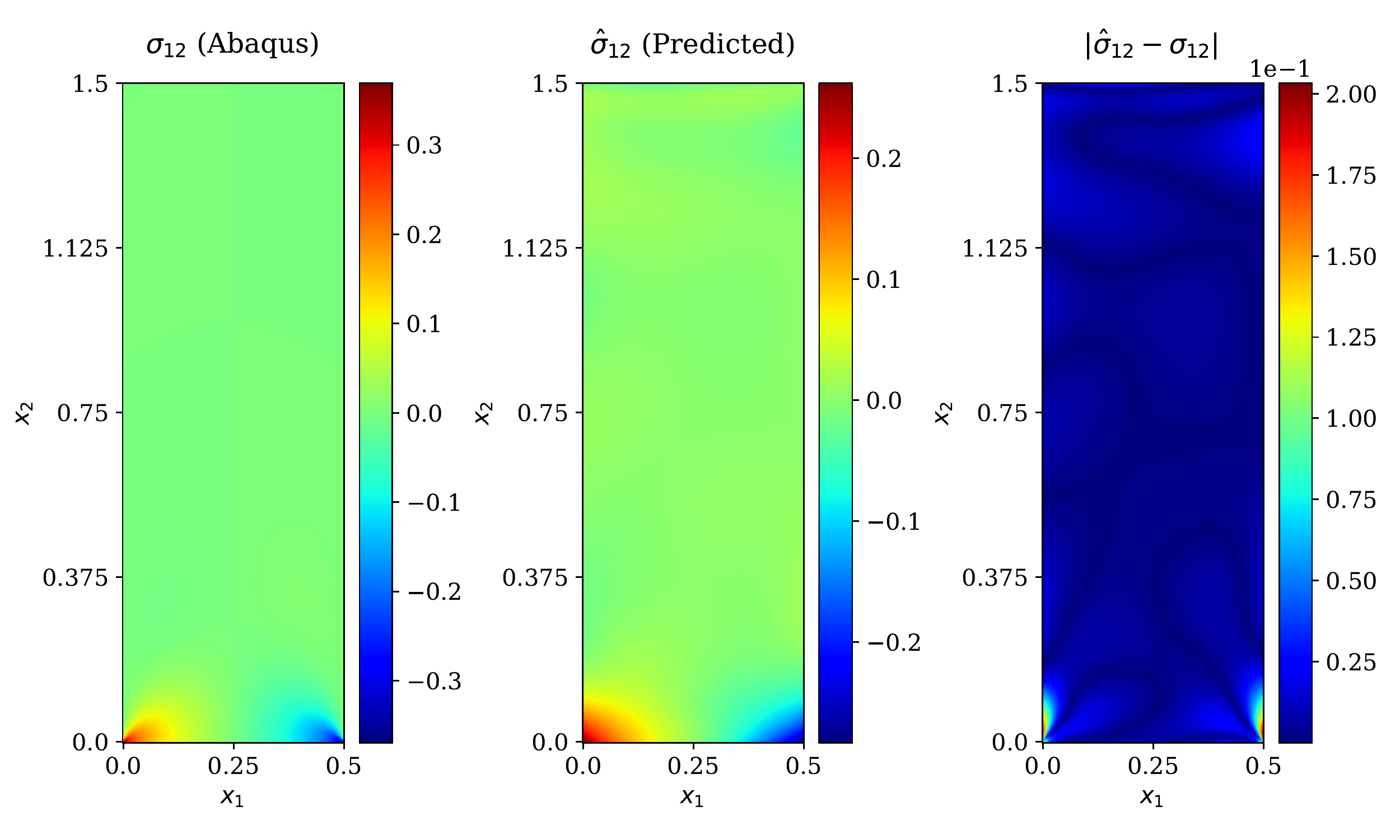}
         \caption{$\sigma_{12}\bkt{\mathbf{x}}$}
     \end{subfigure}
     \hfill
     \begin{subfigure}[b]{\textwidth}
         \centering
         \includegraphics[width=0.7\textwidth]{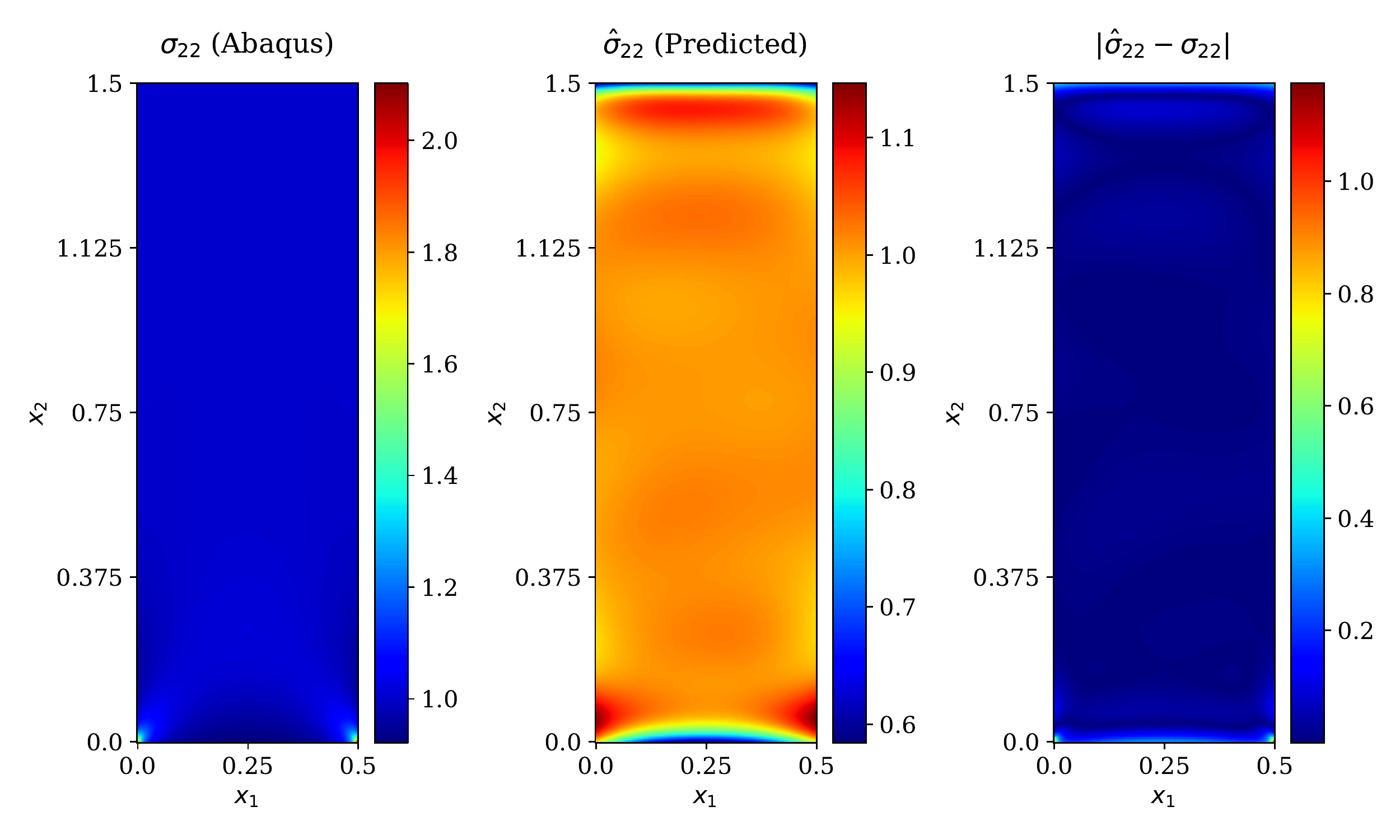}
         \caption{$\sigma_{22}\bkt{\mathbf{x}}$}
     \end{subfigure}
        \caption{Comparison of PINN predicted stress fields and corresponding Abaqus solution for problem defined in section \Cref{sec:2d_fgm_elas_neu}.}
     \label{fig:2d_fgm_elas_neu_stress_comp}
\end{figure}

\noindent A uniform traction $\mathbf{t}$ is applied on the top boundary. Let us say that $ \tilde{v} (\mathbf{x}):\mathrm{\Omega} \to \mathcal{R}^2$ is the neural network model which maps spatial coordinates vector $\mathbf{x}$ in domain $\Omega$ to a real-valued vector in $\mathcal{R}^2$. $\Omega$ represents the set of points that constitute the plate, while $\Gamma$ is the set of points on the boundary of the plate. 
In order to satisfy Dirichlet boundary conditions, the following transformation is applied:
\begin{equation}
    \mathbf{\tilde{u}(x)} = \begin{bmatrix} \tilde{u}_1 \\ \tilde{u}_2 \end{bmatrix} = \begin{bmatrix} x_2 \\ x_2\end{bmatrix} \odot \begin{bmatrix} \tilde{v}_1 \\ \tilde{v}_2 \end{bmatrix}
\end{equation}

\noindent where $u(\mathbf{x})$ gives displacement vector and $\odot$ represents Hadamard product. Decrease in loss (internal strain energy - work done by an external force) with epoch count as the PINN is trained shown in \Cref{fig:neumann_fgm_loss}.
Comparisons between displacement and stress fields predicted by PINN model and the Abaqus FEM simulation are shown in \Cref{fig:neumann_fgm_loss} and \Cref{fig:2d_fgm_elas_neu_disp_comp}, and \ref{fig:2d_fgm_elas_neu_stress_comp}. A good match between the Abaqus and the PINN predicted solution is observed. The R\textsuperscript{2} score, for this case is given in \Cref{tab:r2_scores}. It is to be noted that for $\sigma_{22}$ the R\textsuperscript{2} scores are negative, which usually means quite a bad prediction except when the variance in the target variable is very close to zero. The true value of $\sigma_{22}$ is more or less constant over the domain where R\textsuperscript{2} score is negative. This results in small variance in true target variable and hence a small value in denominator in calculation of R\textsuperscript{2} score (see \Cref{eq:r2_score}). Therefore, even a small absolute error would result in a negative R\textsuperscript{2} score.

\subsection{2D-FGM-ELAS-DIRCH} 
\label{sec:2d_fgm_elas_dirch}
\begin{figure}[h]
    \centering
    \includegraphics[width=80mm]{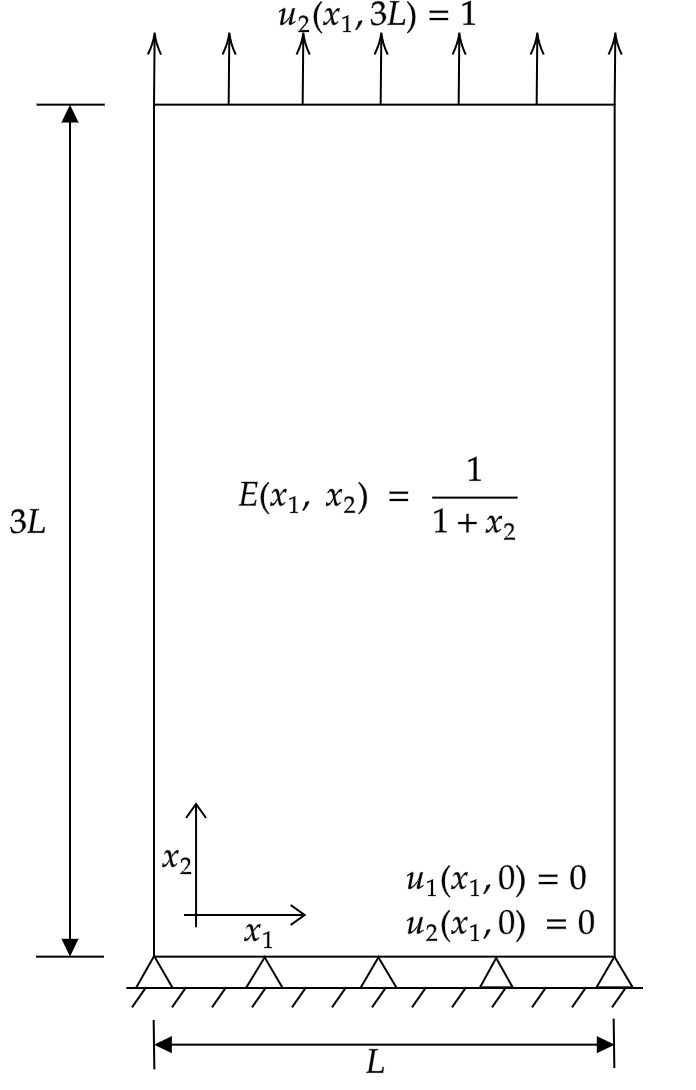}
    \caption{Domain and boundary conditions for functionally graded material object under displacement controlled load defined in \Cref{sec:2d_fgm_elas_dirch}.}
    \label{fig:dirichlet_fgm_2d}
\end{figure}

While we have seen how to train the PINN model on functionally graded material with Neumann boundary conditions, let's consider the same domain and material properties with a different boundary condition. A uniform displacement 
boundary condition is applied on the top edge as shown in the \Cref{fig:dirichlet_fgm_2d}. Since there is no traction, in this case, the loss function will only have the internal strain energy component (see~\Cref{tab:diff_loss_functions}).

\begin{equation} \mathbf{\tilde{u}(x)} = \begin{bmatrix} \tilde{u}_1 \\ \tilde{u}_2 \end{bmatrix} = \begin{bmatrix} x_2\tilde{v}_1 \\ x_2/3L + (3L-x_2)x_2\tilde{v}_2 \end{bmatrix}
\end{equation}

\begin{figure}[h]
    \centering
    \includegraphics[width=80mm]{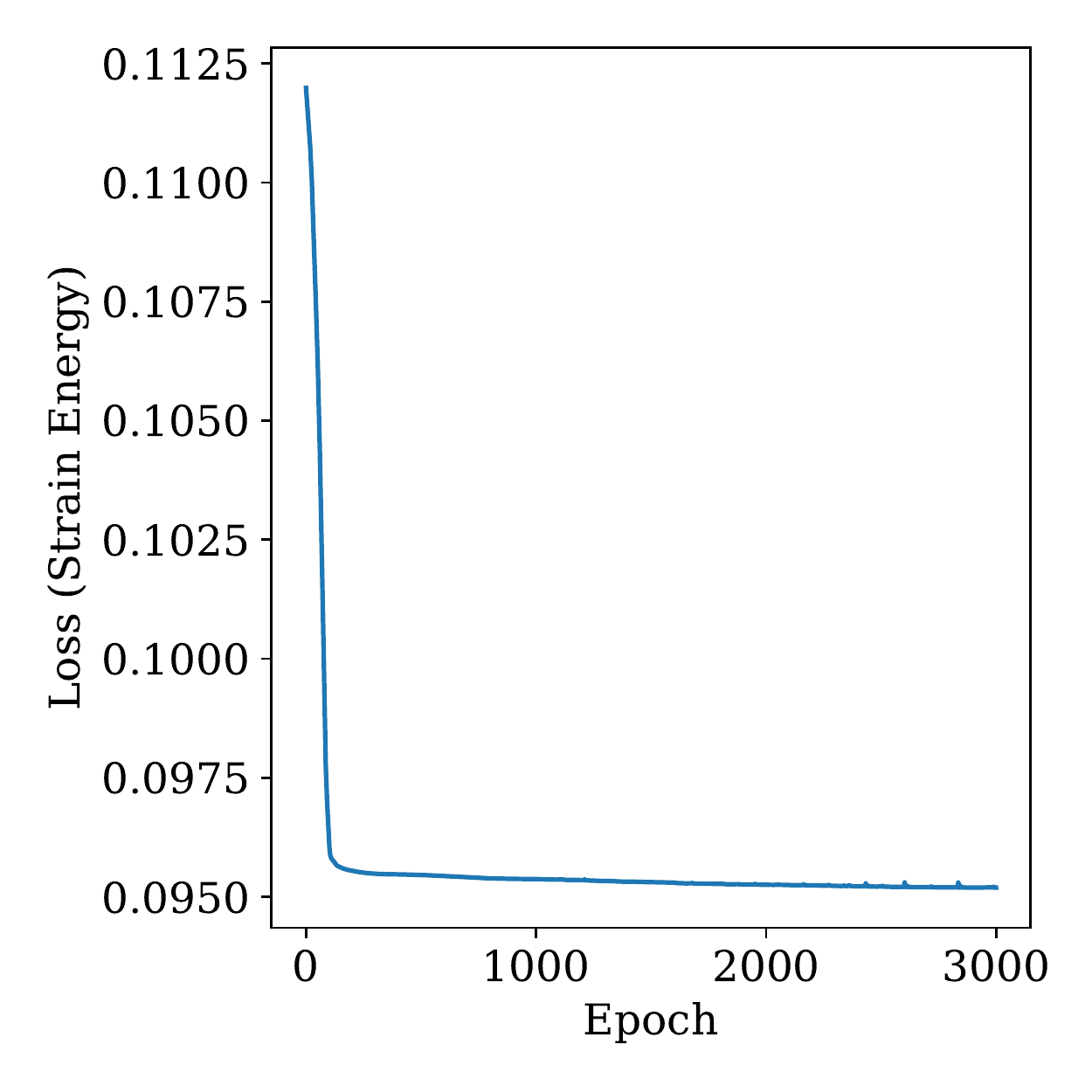}
    \caption{Reduction in loss vs epoch as the PINN is trained to predict elastic response of two-dimensional functionally graded material object with displacement load defined in \Cref{sec:2d_fgm_elas_dirch}.}
    \label{fig:dirichlet_fgm_loss}
\end{figure}

\begin{figure}
     \centering
     \begin{subfigure}[b]{\textwidth}
         \centering
         \includegraphics[width=0.7\textwidth]{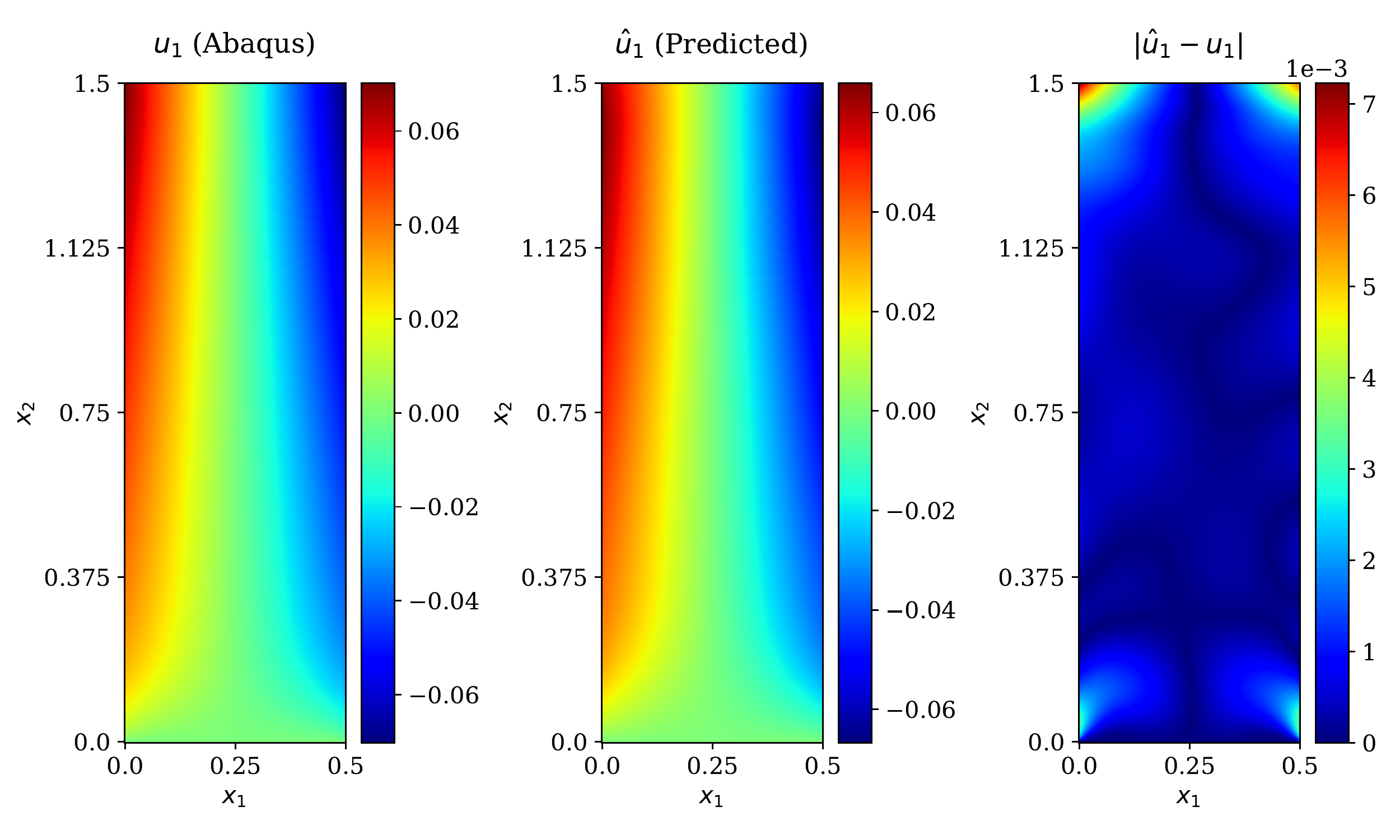}
         \caption{$u_1\bkt{\mathbf{x}}$}
     \end{subfigure}
     \hfill
     \begin{subfigure}[b]{\textwidth}
         \centering
         \includegraphics[width=0.7\textwidth]{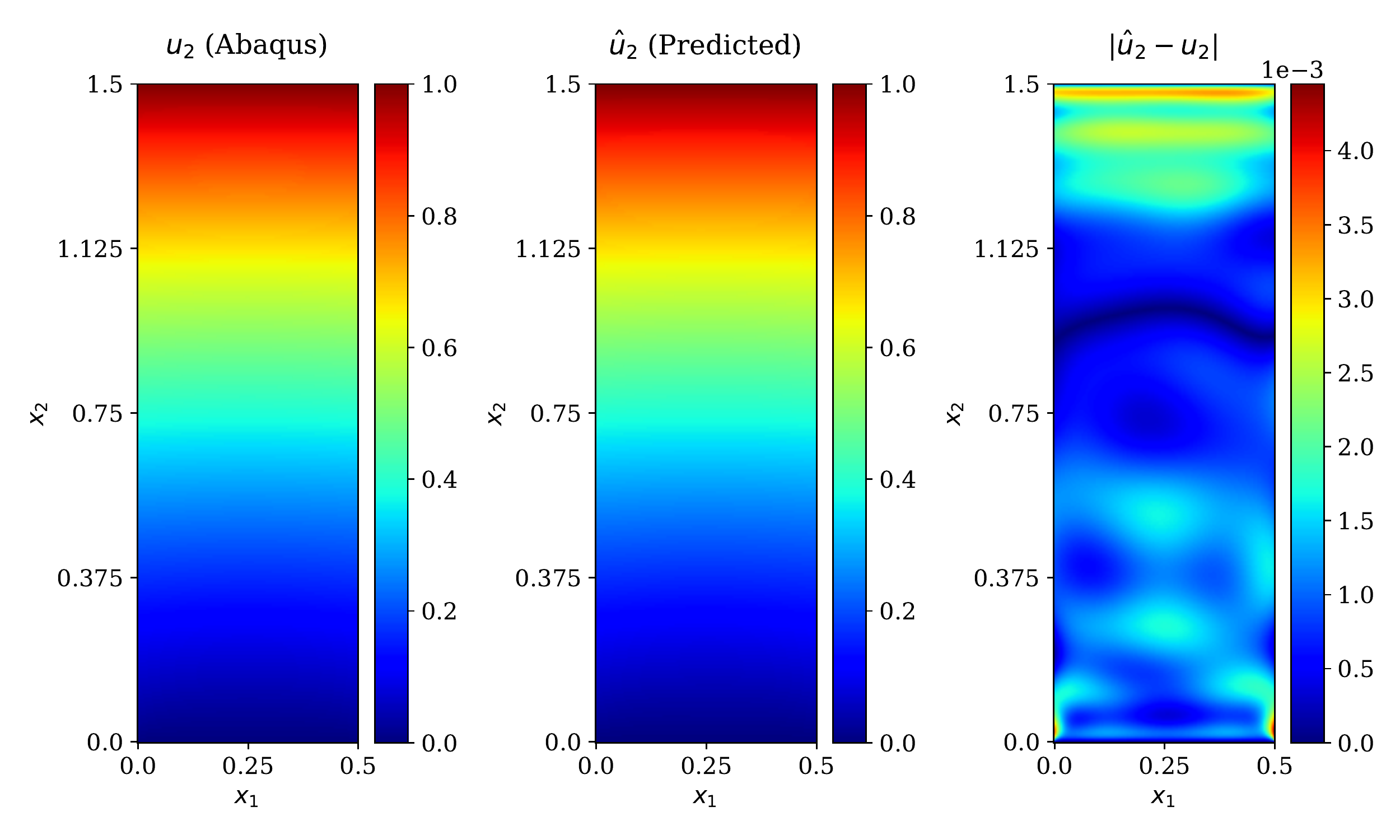}
         \caption{$u_2\bkt{\mathbf{x}}$}
     \end{subfigure}
     \caption{Comparison between PINN prediction displacement fields and corresponding Abaqus solution for problem defined in \Cref{sec:2d_fgm_elas_dirch}.}
     \label{fig:2d_fgm_elas_dirch_disp_comp}
\end{figure}

\begin{figure}
     \centering
     \begin{subfigure}[b]{\textwidth}
         \centering
         \includegraphics[width=0.7\textwidth]{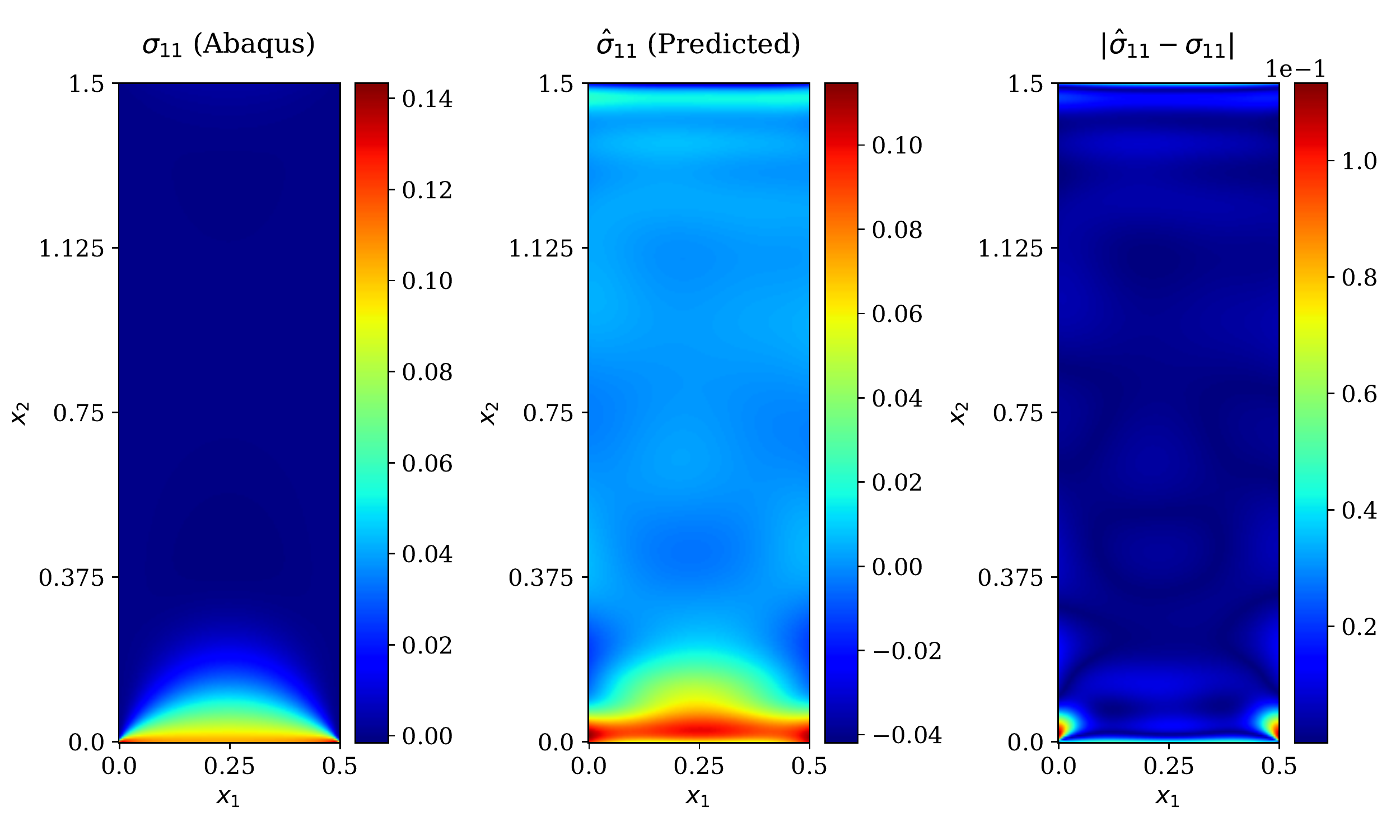}
         \caption{$\sigma_{11}\bkt{\mathbf{x}}$}
     \end{subfigure}
     \hfill
     \begin{subfigure}[b]{\textwidth}
         \centering
         \includegraphics[width=0.7\textwidth]{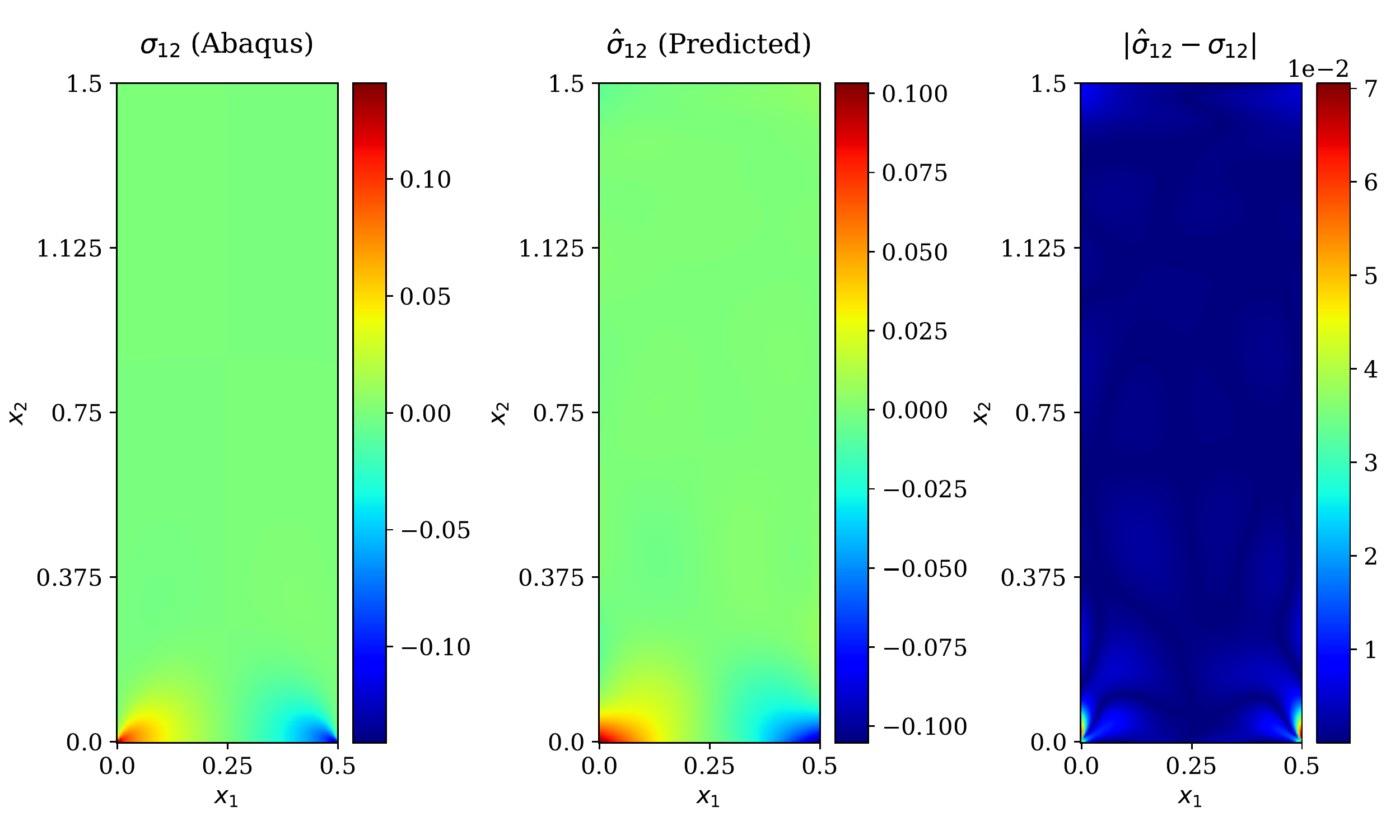}
         \caption{$\sigma_{12}\bkt{\mathbf{x}}$}
     \end{subfigure}
     \hfill
     \begin{subfigure}[b]{\textwidth}
         \centering
         \includegraphics[width=0.7\textwidth]{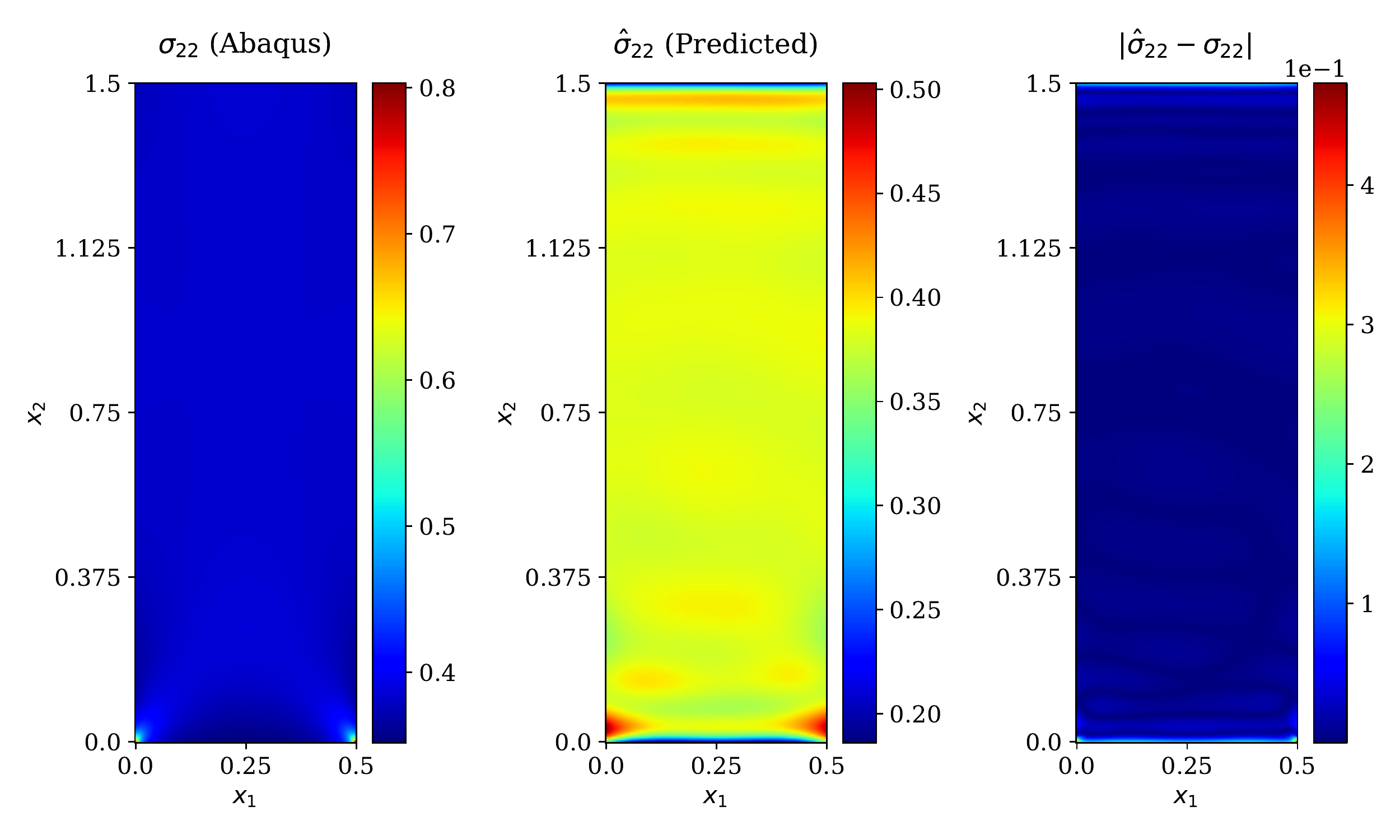}
         \caption{$\sigma_{22}\bkt{\mathbf{x}}$}
     \end{subfigure}
        \caption{Comparison between PINN predicted stress fields and corresponding Abaqus solution for problem defined in \Cref{sec:2d_fgm_elas_dirch}.}
        \label{fig:2d_fgm_elas_dirch_stress_comp}
\end{figure}

Decrease in loss (internal strain energy) with epoch count as the PINN is trained is shown in \Cref{fig:dirichlet_fgm_loss}. Comparisons between stress and displacement fields obtained from PINN and using Abaqus simulation are done in \Cref{fig:2d_fgm_elas_dirch_disp_comp,fig:2d_fgm_elas_dirch_stress_comp}. R\textsuperscript{2} score for this case also was given in \Cref{tab:r2_scores}.

\subsection{2D-FGM-THERMO-ELAS}
\label{sec:2d_fgm_thermo_elas}
\begin{figure}[h]
    \centering
    \includegraphics[width=80mm]{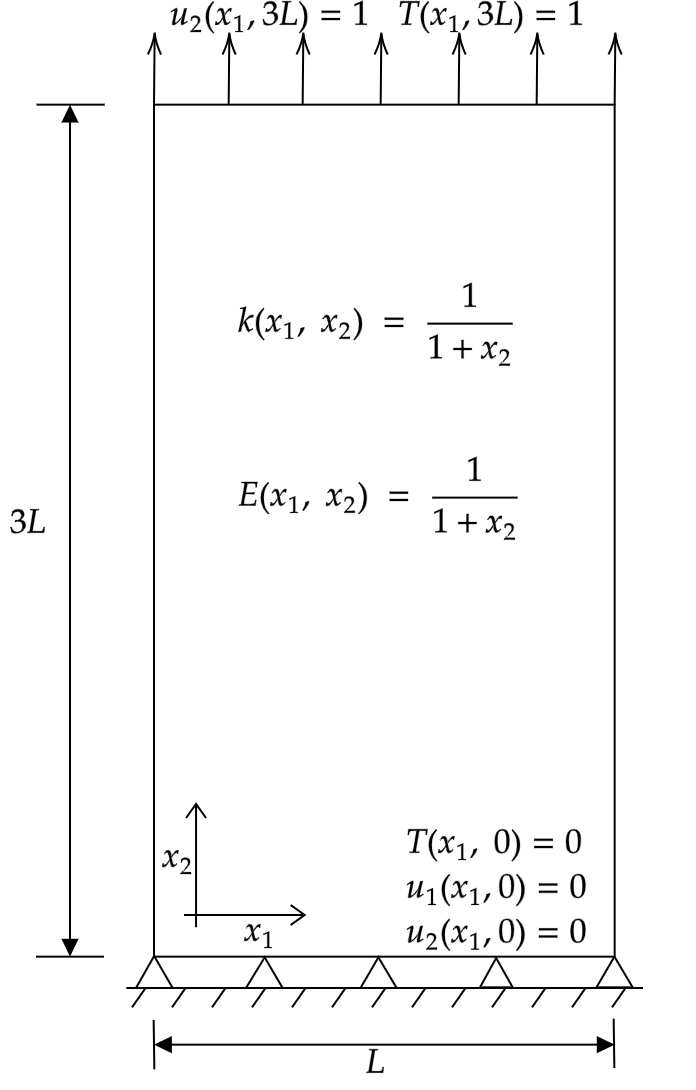}
    \caption{Domain and boundary conditions for two-dimensional functionally graded material object under displacement and temperature load defined in \Cref{sec:2d_fgm_thermo_elas}.}
    \label{fig:thermo_mech_fgm_2d}
\end{figure}

Finally, let us consider a thermo-mechanics problem on 2D domain. The domain and boundary conditions are shown in the \Cref{fig:thermo_mech_fgm_2d}. Both elastic modulus $E(x)$ and conductivity $k(x)$ varies along y-axis such that $ E(x_1, x_2) = {1}/\bkt{1+x_2}$ and $k(x_1, x_2) = {10}/\bkt{1+x_2}$. The coefficient of thermal expansion $\alpha$ is assumed to be unity and the reference temperature $T_0$ is assumed to be zero.

\begin{figure}[h]
    \centering
    \includegraphics[width=90mm]{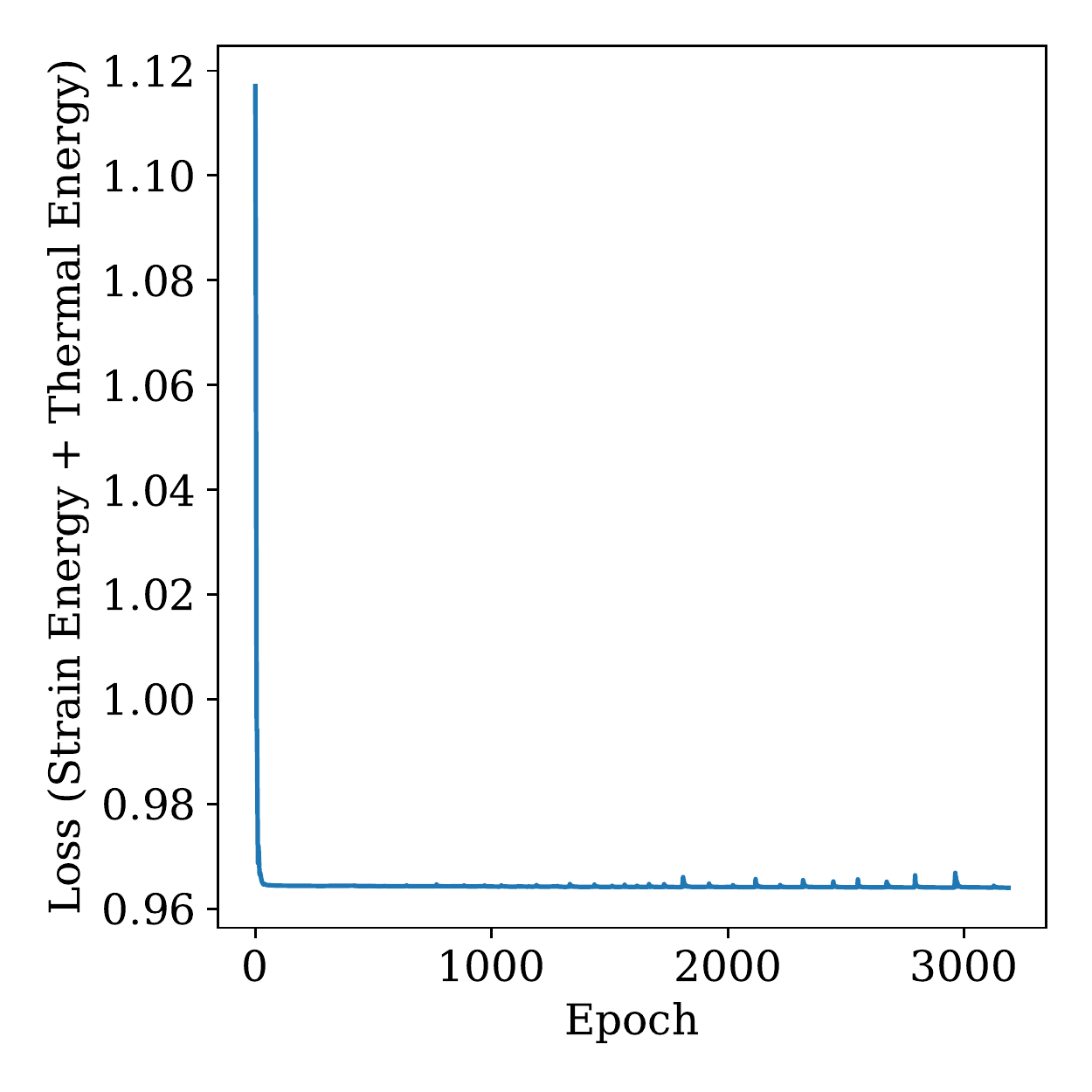}
    \caption{Reduction in internal strain and thermal energy as the model is trained to predict thermo-mechanical response of two-dimensional functionally graded material object defined in \Cref{sec:2d_fgm_thermo_elas}.}
    \label{fig:thermo_fgm_loss}
\end{figure}

\begin{figure}[h]
     \centering
     \begin{subfigure}[b]{\textwidth}
         \centering
         \includegraphics[width=0.7\textwidth]{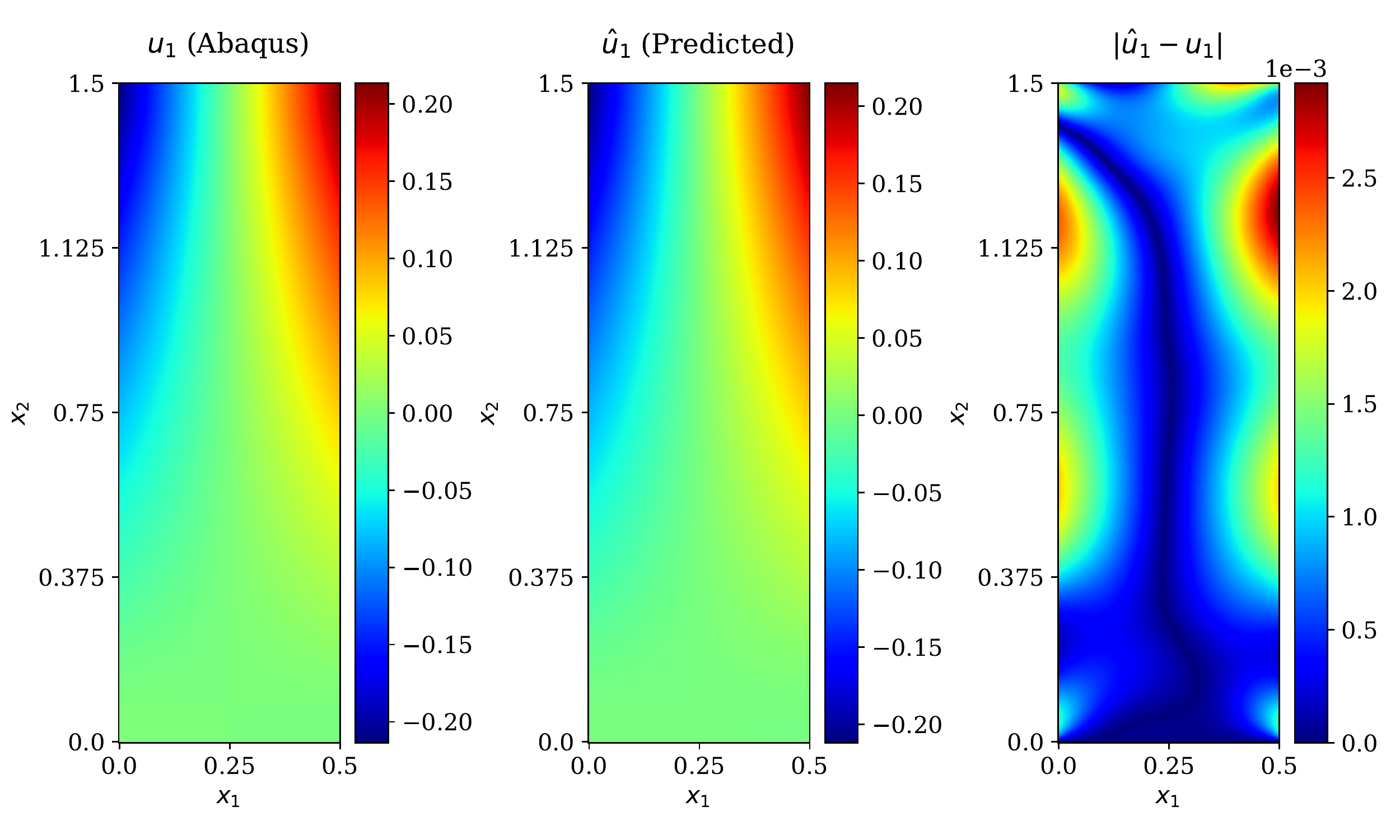}
         \caption{$u_1\bkt{\mathbf{x}}$}
     \end{subfigure}
     \hfill
     \begin{subfigure}[b]{\textwidth}
         \centering
         \includegraphics[width=0.7\textwidth]{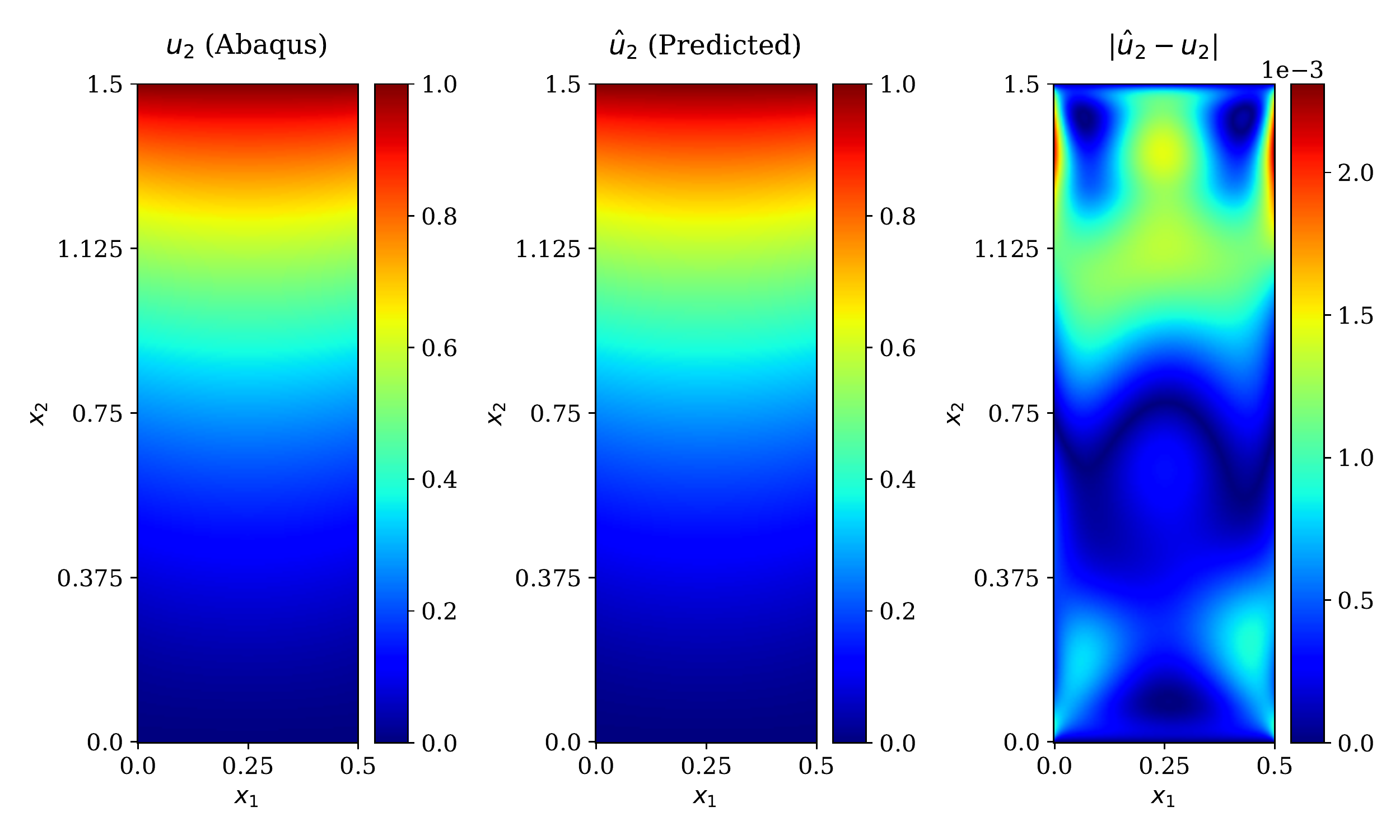}
         \caption{$u_2\bkt{\mathbf{x}}$}
     \end{subfigure}
     \hfill
     \begin{subfigure}[b]{\textwidth}
         \centering
         \includegraphics[width=0.7\textwidth]{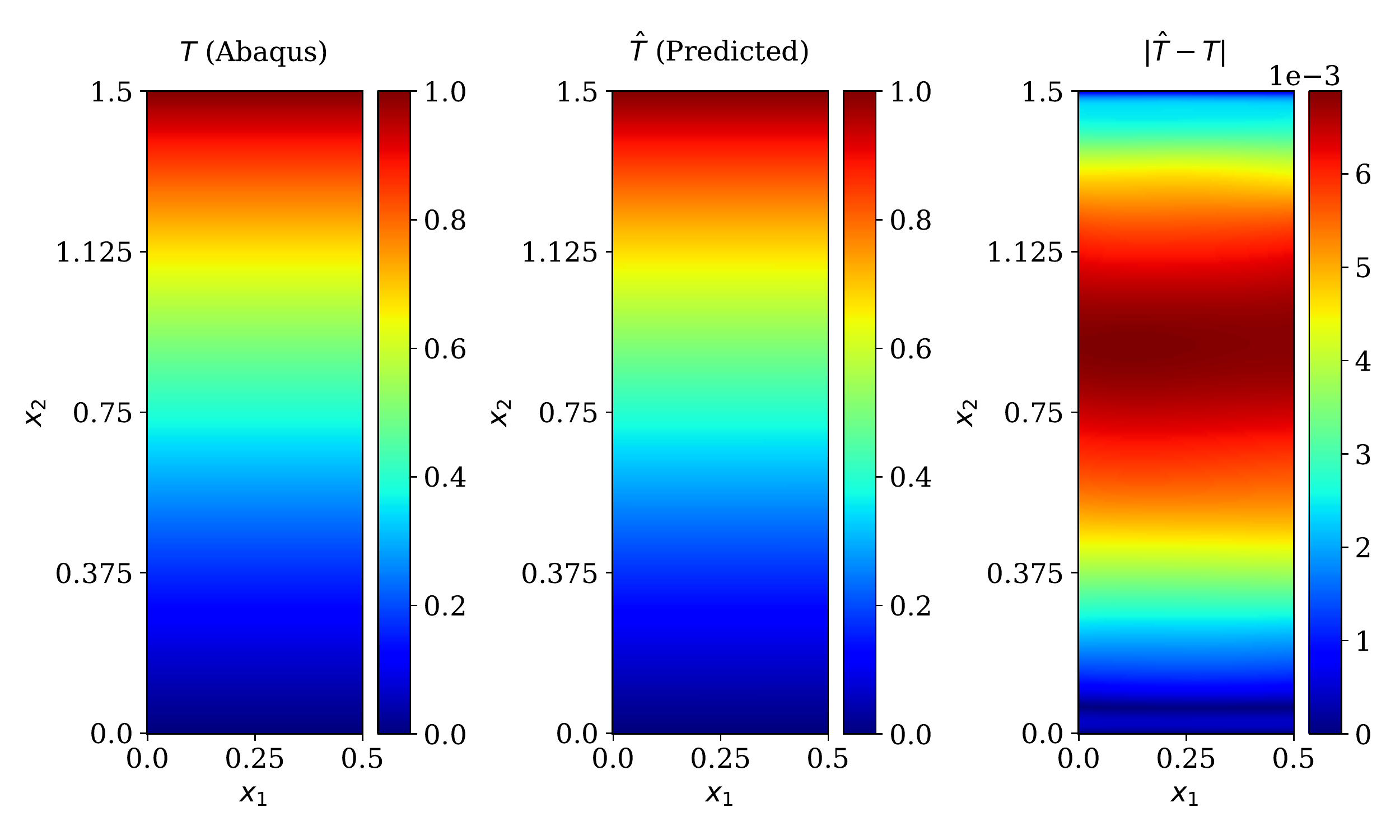}
         \caption{$T\bkt{\mathbf{x}}$}
     \end{subfigure}
     \caption{Comparison between PINN predicted displacement and temperature fields and corresponding Abaqus simulation for problem defined in \Cref{sec:2d_fgm_thermo_elas}.}
     \label{fig:2d_elas_thermo_disp_comp}
\end{figure}

\begin{figure}[h]
     \centering
     \begin{subfigure}[b]{\textwidth}
         \centering
         \includegraphics[width=0.7\textwidth]{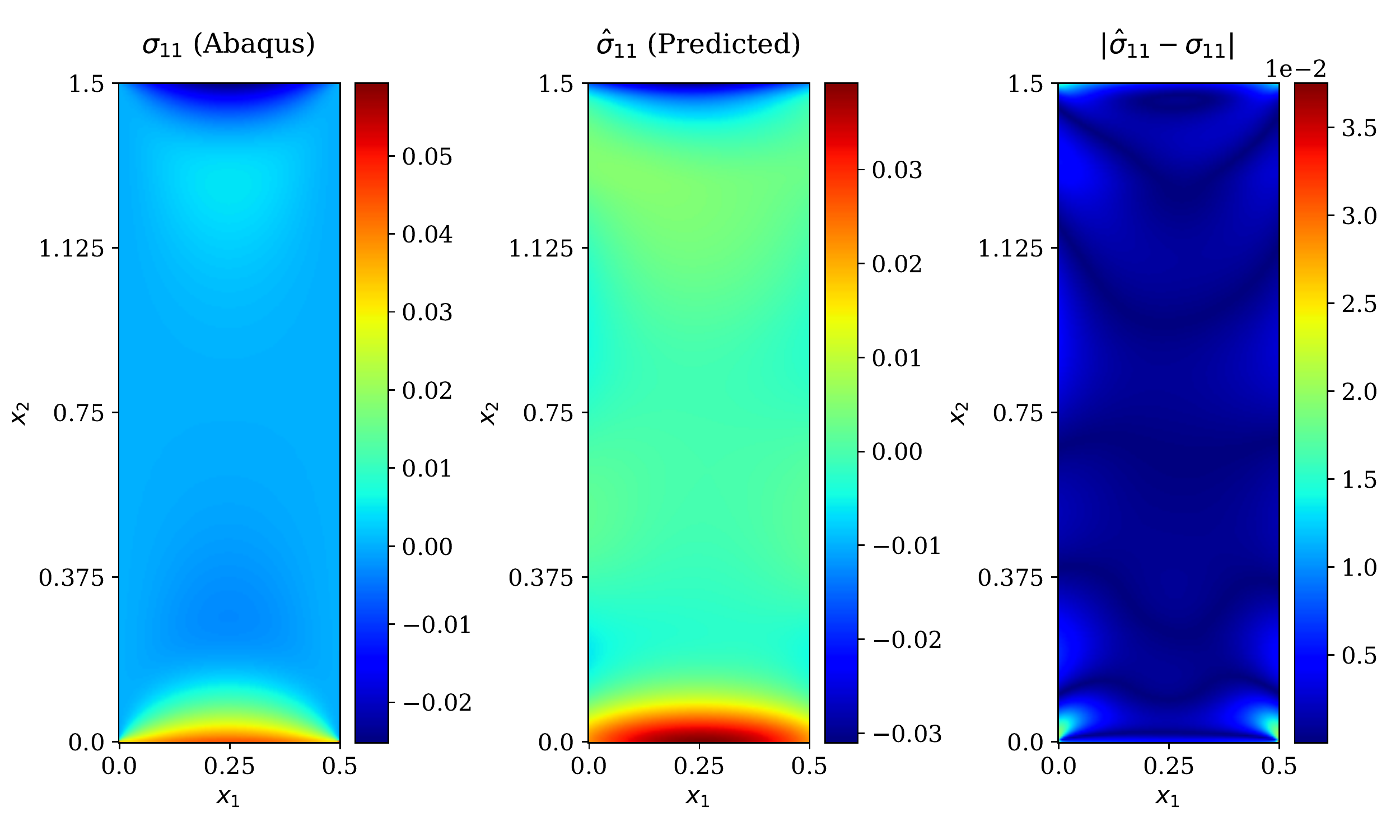}
         \caption{$\sigma_{11}\bkt{\mathbf{x}}$}
     \end{subfigure}
     \hfill
     \begin{subfigure}[b]{\textwidth}
         \centering
         \includegraphics[width=0.7\textwidth]{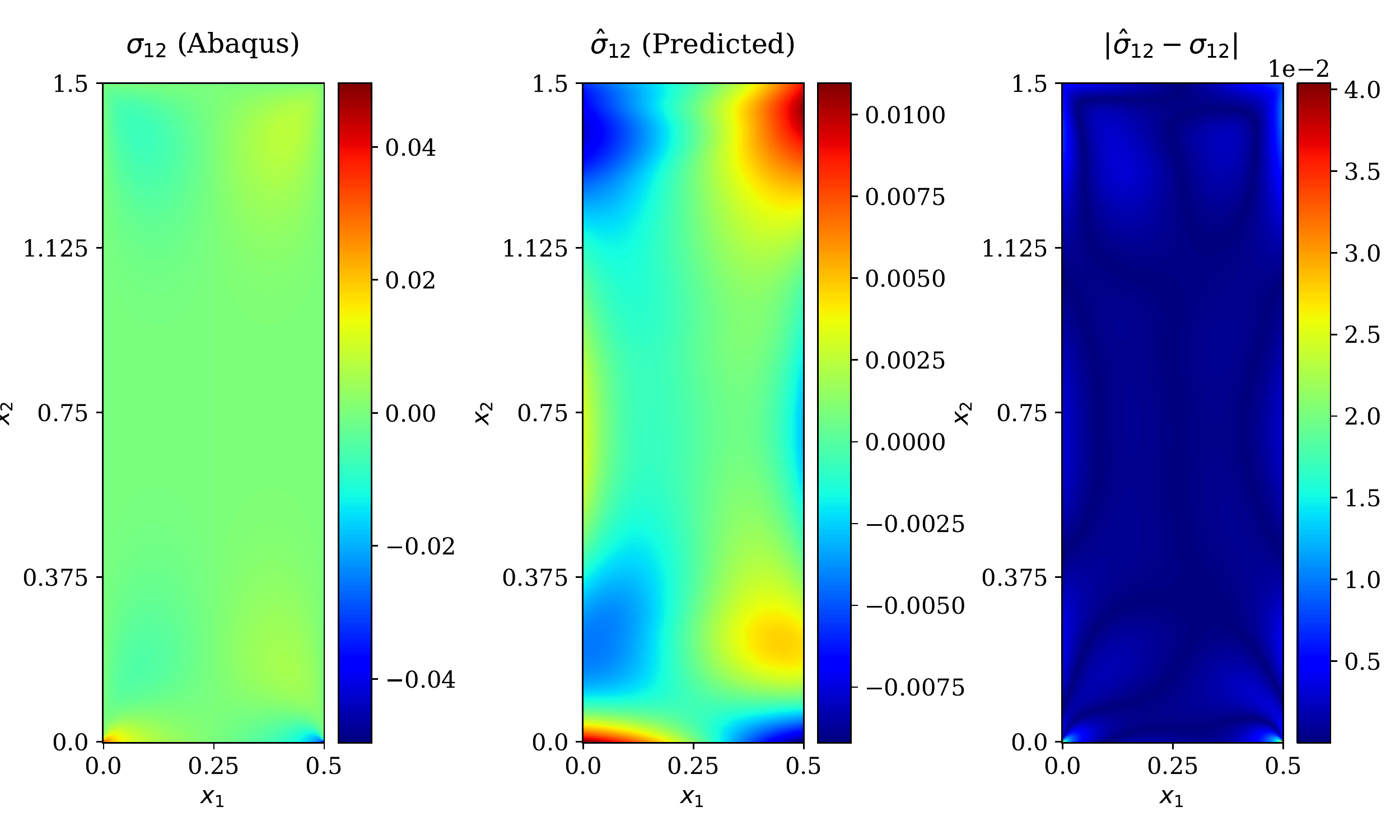}
         \caption{$\sigma_{12}\bkt{\mathbf{x}}$}
     \end{subfigure}
     \hfill
     \begin{subfigure}[b]{\textwidth}
         \centering
         \includegraphics[width=0.7\textwidth]{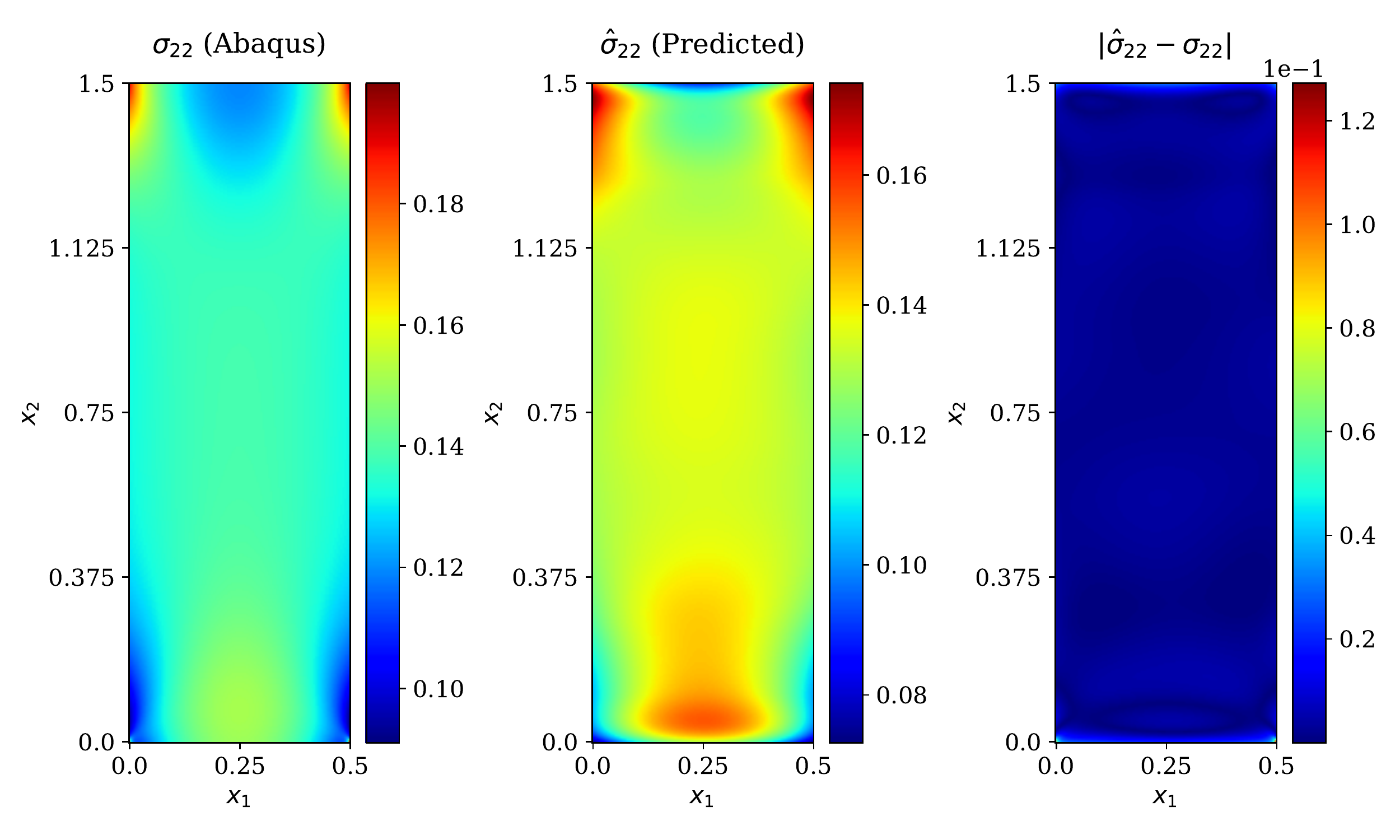}
         \caption{$\sigma_{22}\bkt{\mathbf{x}}$}
     \end{subfigure}
        \caption{Comparision between PINN predicted stress fields and corresponding Aabqus solution for problem defined in \Cref{sec:2d_fgm_thermo_elas}.}
        \label{fig:2d_elas_thermo_stress_comp}
\end{figure}

The differential equations governing the thermo-elastic response of object is given in \Cref{tab:diff_loss_functions}. All primary variables $u_1, u_2$ and $T$ are set to zero on the bottom edge of the plate. The top edge is subjected to vertical displacement of one unit i.e., $u_2(x_2 = 3L) =1$, while temperature of top edge is fixed at one unit $T(x_2 = 3L) = 1$. In order to implement boundary conditions in PINN framework, the blackbox function $\tilde{v}(\mathbf{x})$ is modified as:

\begin{equation}\mathbf{u}\bkt{\mathbf{x}} =
    \begin{bmatrix}
    \tilde{u}_1(\mathbf{x}) \\ \tilde{u}_2(\mathbf{x}) \\ \tilde{T}(\mathbf{x})
    \end{bmatrix} = 
    \begin{bmatrix}
    x_2\tilde{v}_1(\mathbf{x})\\x_2/(3L) + x_2(3L-x_2)\tilde{v}_2(\mathbf{x}) \\ x_2/(3L) + x_2(3L-x_2)\tilde{v}_3(\mathbf{x})
    \end{bmatrix}
\end{equation}
The architecture of neural network $\mathbf{v}\bkt{x}$ is given in \Cref{tab:architecture_details}. The loss function used in training the model is given in \Cref{tab:diff_loss_functions}. The decline in loss as the PINN model is trained can be seen in \Cref{fig:thermo_fgm_loss}.
\noindent After training the model, the PINN predictions are compared with solutions obtained using Abaqus simulation. The comparison between primary fields $u_1(\mathbf{x}), u_2(\mathbf{x}), T(\mathbf{x})$ and stress fields $\sigma_{11}(x), \sigma_{22}(x)$ and $\sigma_{12}(x)$ can be observed in \Cref{fig:2d_elas_thermo_disp_comp} and 
\Cref{fig:2d_elas_thermo_stress_comp}, respectively. The R\textsuperscript{2} score, for different variables is given in \Cref{tab:r2_scores}.

\section{Conclusions}
This study has used PINN to solve solid mechanics and coupled thermo-mechanics problems of objects with functionally graded properties. Through various one and two-dimensional examples, the efficacy of PINNs in accurately predicting primary target variables has been established. On the other hand, it has also been found that PINNs currently do not perform very well in predicting secondary variables such as stress fields. When compared to the finite element simulation method, PINN is advantageous in terms of being meshless and flexible enough that various changes in the form of material properties, domain, and boundary conditions can be incorporated with much ease.

At the same time, there are a few drawbacks of the PINN-based method. There is no knowledge on how to design the neural network given a set of differential equations. The choice of activation function used in PINN influences the kind of functions it can fit. Hence, activation functions play a much significant role in PINN as compared to other neural networks. The PINN framework is also very sensitive to convergence criteria of losses. It is because a small change in loss can sometimes result in a drastic change in the obtained solutions.

The current challenges in using PINN only open the door to further avenues of research. Studying how to design PINN architectures, improving the accuracy of secondary variables, and consistent convergence criteria will be beneficial. Research along the lines of transfer learning using PINN and parallelized training of distributed PINNs will be of great significance.

\section{Appendix}
A neural network is often debugged by looking at various variables such as gradients, activation values, weights, and biases. The evolution of those variables while training the model gives us about model architecture's efficacy. While training PINN, a  slow training speed was observed. We consider the Kirsch's problem discussed earlier in the \Cref{sec:results_and_discussion} and plot the distribution of various variables to get insight into training. It is worthwhile to mention that there are a few open-source libraries such as Tensorboard (\url{https://www.tensorflow.org/tensorboard}), which can help in generating such plots with much ease. Though, in this example, all the plots are generated manually in python.

\begin{figure}[h]
    \centering
    \includegraphics{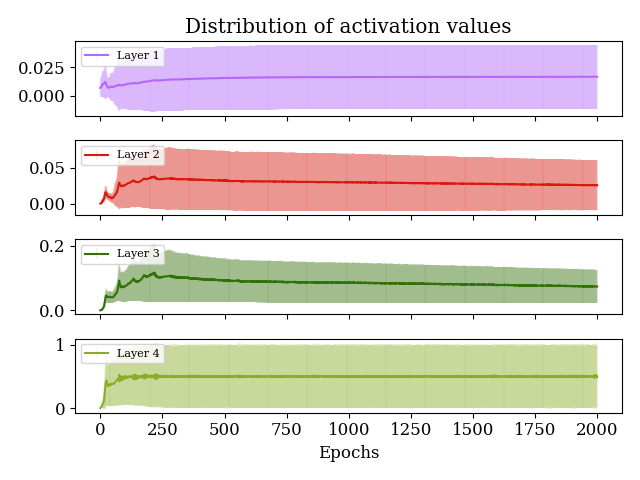}
    \caption{Distribution of activation function values for different layers for PINN trained on problem defined in \Cref{sec:kirschs_problem}. The thick line represents the mean while the faint bars represents the spread about the mean of activation values at various epochs.}
    \label{fig:activation_distribution}
\end{figure}

In \Cref{fig:activation_distribution}, we first look at the distribution of activation values in different hidden layers as the model is trained. The distribution of activation values at the start of the training process would be random due to the random initialization of weights. However, as the training happens, the distribution should be saturated. Moreover, mean activation close to zero with a low standard deviation is not desired as it makes the gradients zero and hampers the training process. In \Cref{fig:activation_distribution}, it is observed that mean of activation values are not zero, which does not impede training. Moreover, the activation values for Layer 2 and Layer 3 have not saturated till 2000 epochs. 

\begin{figure}[h]
    \centering
    \includegraphics{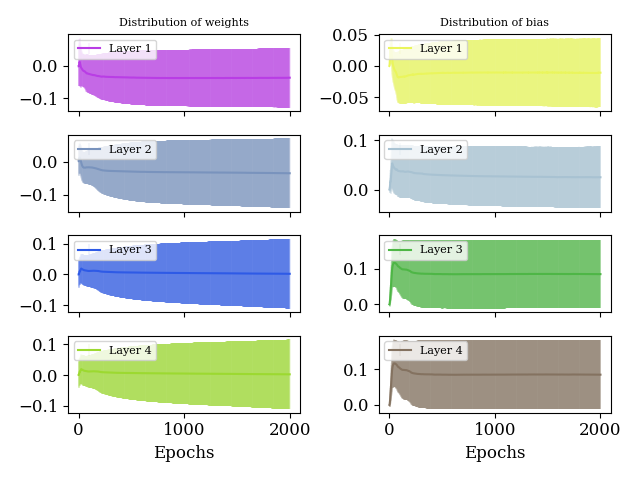}
    \caption{Mean and standard deviation of weights and biases as the PINN is trained for problem defined in \Cref{sec:kirschs_problem}}.
    \label{fig:weight_distribution}
\end{figure}

In \Cref{fig:weight_distribution}, we also look at the means and standard deviations of weights and biases in different layers. We observe that distributions of weights in different layers have not saturated in 2000 epochs as the standard deviation keeps on increasing.

\begin{figure}[h]
    \centering
    \includegraphics{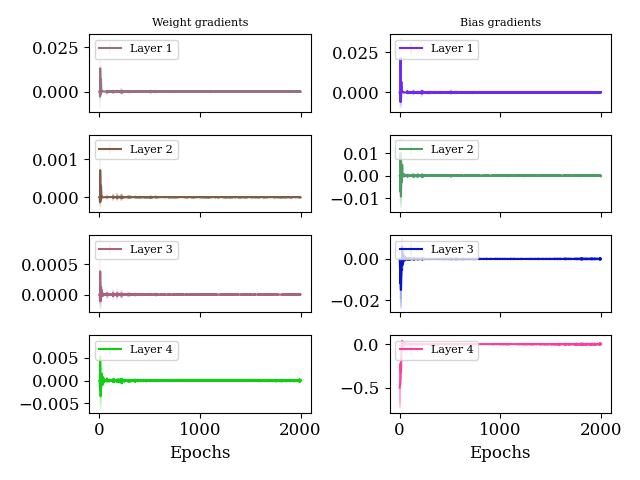}
    \caption{Mean and standard deviation of gradients as the PINN trained for problem defined in \Cref{sec:kirschs_problem}. It is to be noted that zero gradients would result in no further weight updates.}
    \label{fig:gradient_distribution}
\end{figure}

We finally look at \Cref{fig:gradient_distribution}, which shows the distribution of gradients in different layers. We see that the gradients are non-zero at the very start of the training process. Hence, the PINN model trains rapidly for the initial few epochs. However, after a few epochs, the gradients become very close to zero results in slow training of the model. 

\bibliographystyle{elsarticle-num-names}
\bibliography{main}
\end{document}